\documentclass[aps,pra,showpacs,twocolumn,superscriptaddress]{revtex4-1} 
\usepackage{graphicx}
\usepackage{bm}
\usepackage{amssymb} 
\usepackage{amsmath} 
\usepackage{amsthm}
\usepackage{hyperref}
\usepackage[utf8]{inputenc}
\hypersetup{
     colorlinks=true,      
    linkcolor=blue,        
    citecolor=blue,
    filecolor=magenta, 
    urlcolor=black          
}

\usepackage{array, boldline, makecell, booktabs, amsmath}

\providecommand{\ket}[1]{\lvert #1 \rangle}

\providecommand{\ketbra}[2]{\lvert  #1\rangle \langle #2 \rvert}
\newcommand{\expec}[1]{\langle #1\rangle}
\newcommand{\mean}[1]{\big\langle #1\big\rangle}
\newcommand{\meanbig}[1]{\Big\langle #1\Big\rangle}

\usepackage{mathbbol}

\begin{document}
\title{Statistically significant tests of multiparticle quantum correlations  based on randomized measurements}

\author{Andreas Ketterer}
\altaffiliation[Present address: ]{Fraunhofer Institute for Applied Solid State Physics (IAF), Tullastr. 72, 79108 Freiburg, Germany}
\affiliation{Physikalisches Institut, Albert-Ludwigs-Universit\"at Freiburg, Hermann-Herder-Stra\ss e 3,
79104 Freiburg, Germany}
\affiliation{EUCOR Centre for Quantum Science and Quantum Computing, Albert-Ludwigs-Universit\"at Freiburg, Hermann-Herder-Str.~3, 79104~Freiburg, Germany}
\author{Satoya Imai}
\affiliation{Naturwissenschaftlich-Technische Fakult\"at, Universit\"at Siegen, \\Walter-Flex-Stra\ss e 3, 57068 Siegen, Germany}
\author{Nikolai Wyderka}
\affiliation{Institut für Theoretische Physik~III, Heinrich-Heine-Universität Düsseldorf, Universitätsstra\ss e~1, 40225~Düsseldorf, Germany}
\author{Otfried G\"uhne}
\affiliation{Naturwissenschaftlich-Technische Fakult\"at, Universit\"at Siegen, \\Walter-Flex-Stra\ss e 3, 57068 Siegen, Germany}

\begin{abstract}
We consider statistical methods based on finite samples of locally randomized measurements in order to certify different degrees of multiparticle entanglement in intermediate-scale quantum systems. We first introduce hierarchies of multi-qubit criteria, satisfied by states which are separable with respect to partitions of different size, involving only second moments of the underlying probability distribution. Then, we analyze in detail the statistical error of the estimation in experiments and present several approaches for estimating the statistical significance based on large deviation bounds. The latter allows us to characterize the measurement resources required for the certification of multiparticle correlations, as well as to analyze given experimental data in detail. 

\end{abstract}

\maketitle 
%
\section{Introduction}
Noisy intermediate-scale quantum (NISQ) devices involving a few dozen qubits are considered a stepping stone towards the ultimate goal of building a fault-tolerant quantum computer. While impressive achievements have been made in this direction, e.g., in terms of the precision of the individual qubit architectures \cite{TrappedIonRev,QuSup,SupercondQubitsRev}, the common challenge is to scale up the considered devices and, at the same time, maintaining the established accuracy~\cite{FaultTol1,FaultTol2}. In particular, the collective performance of the whole system of interacting qubits is of central concern in this respect. 

Several approaches aiming at a verification of correlation properties of such multiparticle quantum systems have been discussed in the literature \cite{ReviewCertification}. On the one hand, there are efficient protocols in terms of the required measurement resources if the experiment is expected to result in specific states,  e.g., entanglement witnessing~\cite{OtfriedReview}, self-testing~\cite{SelfTestingRev} 
or direct fidelity estimation~\cite{FidelityEst1,FidelityEst2}. 
On the other hand, approaches which rely on few or no expectation  about 
the underlying quantum state are usually very resource-intensive and thus 
do not scale favorably with increasing system sizes, e.g., quantum state tomography~\cite{TomographyReview,CompressedSensing}. Furthermore, intermediate strategies exist which do not aim for a full mathematical description of the 
system but rather focus on specific statistical properties. The latter 
can reduce the required measurement resources considerably at the expense of a non-vanishing statistical error and do not assume any prior information about 
the state~\cite{FlammiaFidelityStat,Shadows1,LeandroRegenerativeModels,RandomMeasTomography,Shadows2}.

\begin{figure}[b!]
\begin{center}
\includegraphics[width=0.42\textwidth]{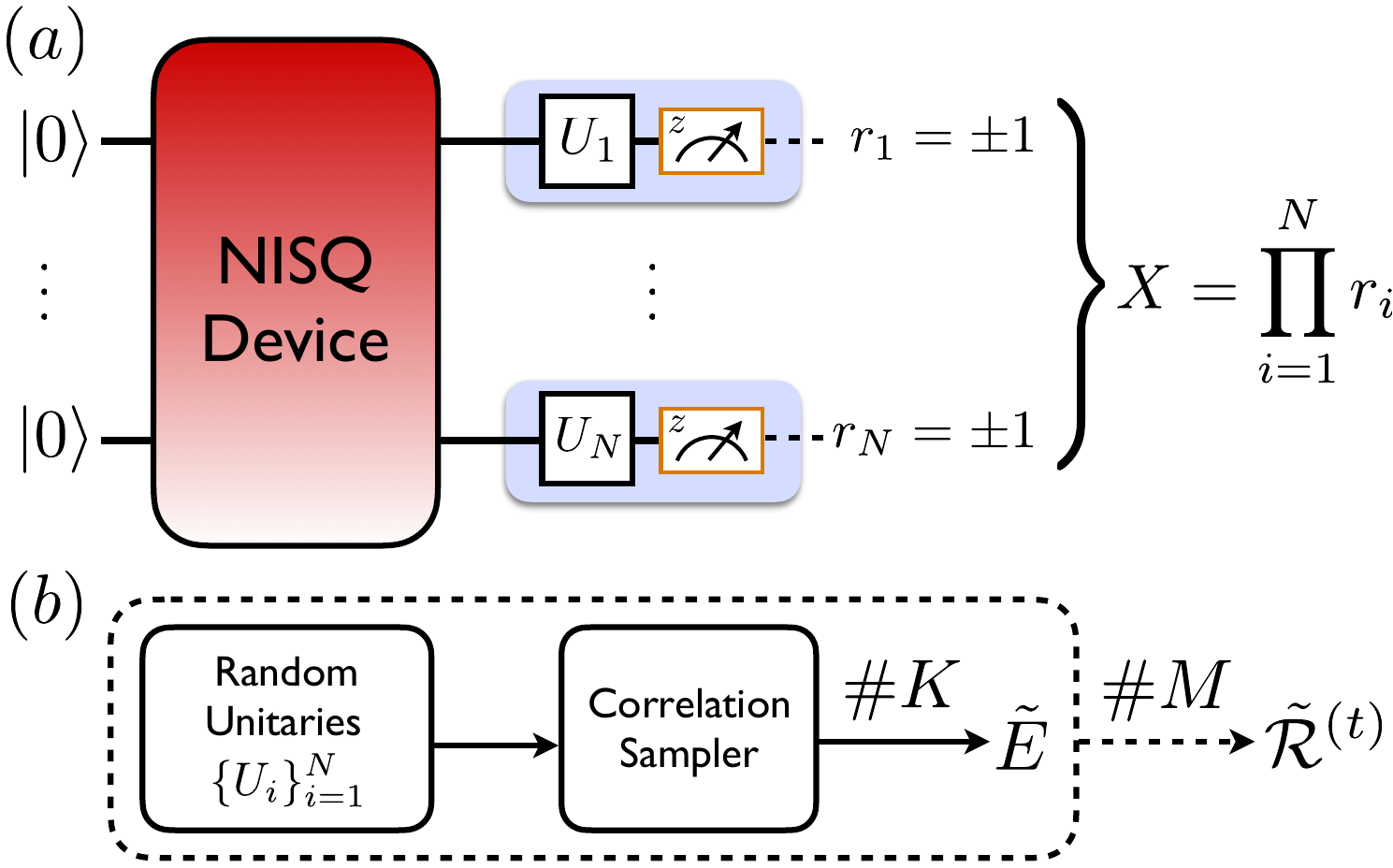}%
\end{center}
\caption{Characterization of a noisy intermediate-scale quantum (NISQ) device through locally randomized measurements. (a) A measurement of $N$ qubits in random local bases defined through the set of local unitary transformations $\{U_i\}_{i=1}^{N}$ resulting in a  correlation sample $X$. (b) Repetition of the measurement protocol presented in (a) for $M$ sets of randomly sample measurement bases and, respectively, $K$ individual projective measurements yields an estimate of the moments~(\ref{eq:RandomMoments}).
}
\label{fig_1}
\end{figure}

Recently there has been much attention on protocols based on statistical correlations between outcomes of randomized measurements \cite{RudolphRandMeasBell,BrunnerRandMeasBell,EnkPRL2012,tran1,tran2,MeMoments1,MeMoments2,MichaelBachelor,TobiasNauck,ZollerFirst,vermerschPRA,ZollerScience,ElbenPRA,DakicRandomMeas,MeineckeExperimentRandom,ElbenPRL,ElbenPRLmixedstate,SatoyaMoments,RandTriads,KnipsPerspective} (see Fig.~\ref{fig_1}). The latter allow to infer several properties of the underlying system, ranging from structures of multiparticle entanglement~\cite{MeMoments1,MeMoments2,SatoyaMoments}, over subsystem purities~\cite{ZollerScience,ElbenPRA}, to fidelities with respect to certain target states or even another quantum devices \cite{FlammiaFidelityStat,ElbenPRL}. At the core of all those approaches is the idea to perform measurements in 
randomly sampled local bases leading to ensembles of measurement outcomes 
whose distributions provide a fingerprint of the system's correlation 
properties. Concerning resources required for statistically significant
tests, scaling properties have been derived for the case of bipartite
entanglement \cite{vermerschPRA, ElbenPRLmixedstate}.

In this work we present detailed statistical methods to 
certify multiparticle entanglement structures in systems consisting of many qubits. First, we derive criteria in 
terms of second moments of randomized measurements for different forms of multiparticle entanglement allowing to infer the entanglement 
depth. Second, we present several rigorous approaches for the analysis of the underlying statistical errors, based on large deviation bounds, which are of great relevance for practical experiments.  
As we will see, our results may directly be used in
current experiments using Rydberg atom arrays or superconducting qubits
\cite{20qExpGHZ1, 20qExpGHZ2a, 20qExpGHZ2}.

\section{Moments of random correlations}
We consider a mixed quantum state of $N$ qubits described by the density matrix $\rho$. 
In order to characterize this state we follow a strategy based on locally randomized measurements.  Each random measurement is characterized through a set of random bases $\{(\ket{u_n^{(0)}}= U_n\ket{0_n},\ket{u_n^{(1)}}= U_n\ket{1_n})\}_{n=1,\ldots,N}$, with  $\{U_n\}_{n=1,\ldots,N}$ picked from the unitary group $\mathcal U(2)$ according to the Haar measure. Further on, we can  associate to each element $(\ket{u_n^{(0)}}= U_n\ket{0_n},\ket{u_n^{(1)}}= U_n\ket{1_n})$, with $n\in \{1,\ldots,N\}$, a direction $\boldsymbol u_n$ on the unit sphere $S^{2}$ with components $[\boldsymbol u_n]_i=\mathrm{tr}[\sigma_{\boldsymbol u_n}\sigma_i]$, with $i\in \{x,y,z\}$, and $\sigma_{\boldsymbol u_n}= U_n \sigma_z U_n^\dagger$ (see Fig.~\ref{fig_1}(a)). One such random measurement then leads to the correlation function:
\begin{align}
E(\boldsymbol u_1,\ldots,\boldsymbol u_{N})=\expec{ \sigma_{\boldsymbol u_1}\otimes \ldots  \otimes \sigma_{\boldsymbol u_{N}}}_{\rho},
\label{eq:CorrelationFct}
\end{align}
which provides a random snapshot of the correlation properties of the output state $\rho$. In order to get a more complete picture we consider the corresponding moments
\begin{align}
\mathcal R^{(t)}= &\frac{1}{(4\pi)^{N}} \int_{S^{2}} d\boldsymbol u_1\ldots \int_{S^{2}} d\boldsymbol u_{N} 
\left[E(\boldsymbol u_1,\ldots,\boldsymbol u_{N})\right]^t, 
\label{eq:RandomMoments}
\end{align}
where $t$ is a positive integer and $d\boldsymbol u_i=\sin{\theta_i} d\theta_i d\phi_i$ denotes the uniform measure on the sphere $S^{2}$. The moments (\ref{eq:RandomMoments}) are by definition invariant under local unitary transformation
 and thus good candidates for the characterization of multiparticle correlations. 

\section{Multiparticle entanglement characterization}
In a multiparticle system one defines $k$-separable states, with $k\in \{2,\ldots,N\}$, as those states which can be written as a statistical mixture of $k$-fold product states $\ket{\Psi^{(k)}}=\ket{\phi_1}\otimes \ldots \otimes \ket{\phi_k}$. Hence, by disproving that a state belongs to the above separability classes one can infer different degrees of multiparticle entanglement, with the strongest form given by states which are not even $2$-separable, i.e., genuinely multiparticle entangled (GME). The concept of $k$-separability is a widely used approach to benchmark experiments~\cite{BenchAppl1,BenchAppl2,BenchAppl3,BenchAppl4} and also has been identified as resource in quantum metrology applications~\cite{MetroAppl1,MetroAppl2,MetroAppl3,MetroAppl4}. 

To begin with we note the well-known criterion $\mathcal R^{(2)}\leq 1/3^N$ which holds for all $N$-separable (i.e. fully-separable) states~\cite{tran1,tran2,BriegelLUinv,JulioMarkus,ZukowskiRefFrame1,ZukowskiRefFrame2}. Furthermore, bi-separability bounds on combinations of second moments of marginals of three-qubit systems can be formulated~\cite{MintertPRL2005,SatoyaMoments}. However, so far no useful bounds on the full $N$-qubit moments~(\ref{eq:RandomMoments}) for the detection of GME have been found. Here we close this gap and prove in App.~\ref{app:ProofMainCriteria} that all $k$-separable mixed states fulfill the bounds
\begin{align}
 \mathcal R^{(2)}\leq\frac{1}{3^{N-k+1}}\times \begin{cases}
2^{N-(2k-1)}, &N\ \mathrm{odd}, \\
2^{N-(2k-1)}+1,  &N\ \mathrm{even},
\end{cases}
\label{eq:ksepcrit}
\end{align}
with $k=2,\ldots,\left \lfloor{{(N-1)}/2}\right \rfloor $. Equation~(\ref{eq:ksepcrit}) thus provides a hierarchy of entanglement 
criteria whose violation for fixed $k$ implies that the given state is 
at most $(k-1)$-separable. This implies that it has an entanglement depth 
\cite{sorensenprl, guehnenjp} of at least $\lceil{N/(k-1)}\rceil$,
but possibly stronger bounds for the depth can be derived based on the concept of producibility, see App.~\ref{app:EntDepth}, Sec.~I.E. In any case, only states which are GME can reach the maximum value of the second moment $\mathcal R^{(2)}$ which is known to be attained by the 
$N$-qubit GHZ states~\cite{JensR2max,NikolaiSectorLengths}:
\begin{align}
 \mathcal R^{(2)}_{\ket{\text{GHZ}_N}}=\frac{1}{3^N}\times \begin{cases}
2^{N-1}, &N\ \mathrm{odd}, \\
2^{N-1}+1,  &N\ \mathrm{even}.
\end{cases}
\label{eq:GHZsts}
\end{align}
with $\ket{\text{GHZ}_N}=(\ket 0^{\otimes N}+\ket 1^{\otimes N})/\sqrt 2$. Note that in systems consisting of larger local dimensions it is in general not true that states which maximize the corresponding generalized second moment $\mathcal R^{(2)}$ are GME~\cite{JensR2max,NikolaiSectorLengths,SatoyaMoments}.

\begin{figure}[t!]
\begin{center}
\includegraphics[width=0.42\textwidth]{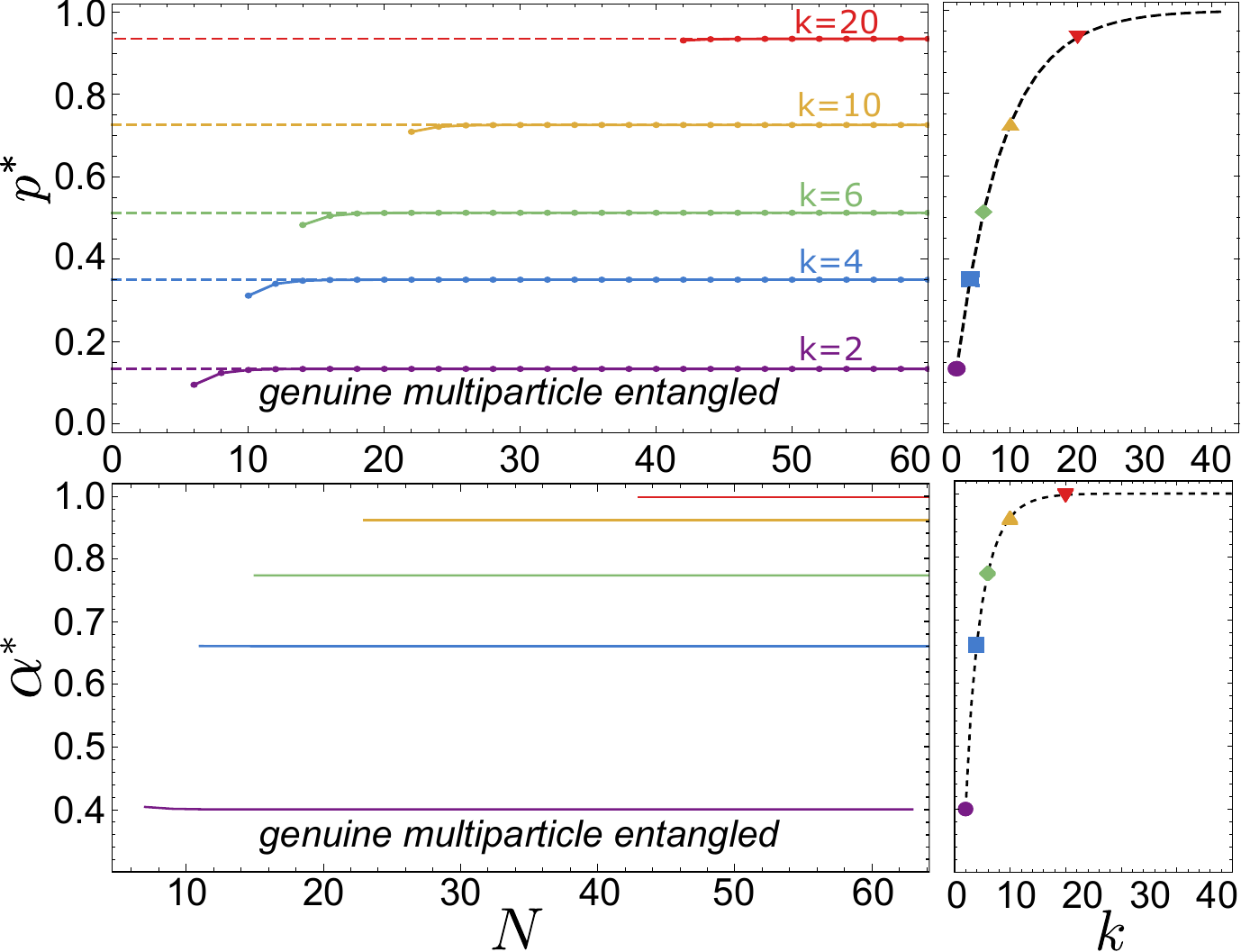}%
\end{center}
\caption{Threshold values $p^*$ (top) and $\alpha^*$ (bottom) up to which the noisy  and the asymmetric GHZ-state, $\rho_\text{GHZ}^{(N)}(p)$ and $\ket{\text{GHZ}^{(N)}_\alpha}$, respectively, are detected to be not $2$- (violet, bottom), $4$-  (blue), $6$- (green), $10$- (yellow) and $20$-separable (red, top) as a function of the number of qubits $N$. In the upper panel dots connected by solid lines represent values of $p^*$ for even $N$, dashed lines correspond to the case of odd $N$. Plots in the right collumn show the asymptotic values of $p^*$ and respectively $\alpha^*$ in the limit $N\rightarrow \infty$  as a function of the parameter $k$. The exemplary values corresponding to the left plot are highlighted by colored markers. 
}
\label{fig_2}
\end{figure}

In the following we study the performance of the criteria~(\ref{eq:ksepcrit}) by considering the noisy $N$-qubit GHZ states $\rho_\text{GHZ}^{(N)}(p):=p \mathbb 1/2^N+(1-p) \ketbra{\text{GHZ}_N}{\text{GHZ}_N}$, which yields $\mathcal R_\text{GHZ}^{(2)}(p,N)=(1-p)^2  \mathcal R^{(2)}_{\ket{\text{GHZ}_N}}$ and thus minimizes (maximizes) $\mathcal R^{(2)}$ for $p=1$ ($p=0$). We note that in practical situations where errors occur locally one can estimate the global depolarization probability $p$ by combining local depolarization rates corresponding for instance to average gate errors (see App.~\ref{app:OtherNoiseSources}). The threshold value of $p$ up to which (\ref{eq:ksepcrit}) is violated as a function of $N$ and $k$ thus reads
\begin{align}
p^*=1-f(N,k)\left({\frac{3}{4}}\right)^{\frac{k-1}{2}} ,
\label{eq:threshold}
\end{align}
where $f(N,k)=1$ for odd $N$ and $f(N,k)=\sqrt{(4^k+2^{N+1})/(4+2^{N+1})}$ for even $N$ (see  Fig.~\ref{fig_2}). As is clear from Eq.~(\ref{eq:threshold}), the threshold $p^*$ is independent of $N$, for odd $N$, and coincides with the asymptotic threshold in the limit $N\rightarrow \infty$, where $f(N,k)\rightarrow 1$. 
The latter is strictly smaller than $1$ which shows that Eq.~(\ref{eq:ksepcrit}) can be applied also in systems consisting of a large number of parties. Furthermore, our methods also work in the regime of low fidelities, i.e., large $p^*$'s, where fidelity-based witnesses fail (see Fig.~\ref{fig_2}). Furthermore, in the lower panel of Fig.~\ref{fig_2} we analyse the performance of the criteria~(\ref{eq:ksepcrit}) for GHZ states with unequal amplitudes, i.e., $\ket{\text{GHZ}^{(N)}_\alpha}=\sqrt{(1+\alpha)/2} \ket 0^{\otimes N}+\sqrt{(1-\alpha)/2} \ket 1^{\otimes N}$, with $0\leq\alpha \leq 1$ (see also App.~\ref{app:BoundsMoments}). 
Lastly, we note that the criteria (\ref{eq:ksepcrit}) become useful only for a certain minimum number of qubits depending on the value of $k$, e.g., GME detection is only possible for $N>4$.

%
\section{Estimation of the moments}
In the following we assume that a finite sample of $M$ random measurement bases is taken, each of which undergoes $K$ individual projective measurements.  
We thus denote the outcomes of a single random measurement on $N$ qubits by $\{r_1,\ldots,r_{N}\}$, with $r_i=\pm1$, and define the corresponding correlation sample as $X=\prod_{i=1}^{N}r_i$ (see Fig.~\ref{fig_1}(a)). Given a fixed measurement basis we can thus model the binary outcomes of $X$ through a binomially distributed random variable $\tilde Y$ with probability $P$, i.e., the probability that an even number of the measurement outcomes $r_i$ result in $-1$, and $K$ trials. The corresponding unbiased estimators $\tilde P_k$ of $P$ and its $k$-th powers, respectively, are then given by  $\tilde P_k=\tilde P_{k-1}[K\tilde P_1-(k-1)]/[K-(k-1)]$, with $\tilde P_1=\tilde Y/K$ (see App.~\ref{app:Estimators}).
\\
\indent Further on, the unbiased estimators of the respective $t$-th powers of Eq.~(\ref{eq:CorrelationFct}) read
\begin{align}
\tilde E_t=(-1)^t\sum_{k=0}^t (-2)^k {t\choose k} \tilde P_k
\label{eq:EstimatorEt}
\end{align}
which, in turn, allows us to define faithful estimators of the moments~(\ref{eq:RandomMoments}), resulting from $M$ sampled measurement bases:
\begin{align}
\tilde{\mathcal R}^{(t)}=\frac{1}{M}\sum_{i=1}^M  [\tilde E_t]_i.
\label{eq:EstimatorRt}
\end{align}
Given Eqs.~(\ref{eq:EstimatorEt}) and (\ref{eq:EstimatorRt}), our goal is now to gauge the statistical error of an estimation $\tilde{\mathcal R}^{(t)}$ as a function of the number of subsystems $N$. More precisely, we aim for lower bounds on the total number of required measurement samples $M_\text{tot}=M\times K$ needed in order to estimate $\mathcal R^{(t)}$ with a precision of at least $\delta$ and confidence $\gamma$ , i.e.,  such that $\text{Prob}(|\tilde{\mathcal R}^{(t)}-{\mathcal R}^{(t)}|\leq \delta)\geq \gamma $ for $M_\text{tot}\geq M(\gamma,t)$. 
\\
\indent
In order to achieve this goal we exploit concentration inequalities which provide deviation bounds on the probability $1-\text{Prob}(|\tilde{\mathcal R}^{(t)}-{\mathcal R}^{(t)}|\leq \delta)$, i.e., the probability that the estimator deviates from the mean value by a certain margin. In App.~\ref{app:EstimationMoments} we discuss three such approaches which differ in their assumptions on the random variable $\tilde{\mathcal R}^{(t)}$, based on the Chebyshev-Cantelli and Bernstein inequality, as well as a more general approach using Chernoff bounds~\cite{SchmidtSpringer2010}. For instance, for the Chebyshev-Cantelli inequality this leads to a minimal two-sided error bar of $\tilde{\mathcal R}^{(t)}$ that guarantees the confidence $\gamma$:  
\begin{align}
\delta_\text{err}(\gamma)= \sqrt{\frac{1+\gamma }{1-\gamma } \text{Var}{\left(\tilde{\mathcal R}^{(t)}\right)}},
\label{eq:Precision1}
\end{align}
where $\text{Var}(\tilde{\mathcal R}^{(t)})$ denotes the variance of the estimator~(\ref{eq:EstimatorRt}) which can be evaluated using  the properties of the binomial distribution. For instance, in case of the second moment $\mathcal R^{(2)}$ we find that the variance reads
\begin{align}
\text{Var}(\tilde{\mathcal R}^{(2)})=&\frac{1}{M} \left[A(K)\mathcal R^{(4)}+B(K)\mathcal R^{(2)}\right. \nonumber \\
&\left. +C(K)-\left(\mathcal R^{(2)}\right)^2\right],
\label{eq:VarianceR2tilde}
\end{align}
with $A(K)=(K-2)(K-1)C(K)/2$, $B(K)=2(K-2)C(K)$ and $C(K)=2/[K(K-1)]$ which are determined through the properties of the binomial distribution (see App.~\ref{app:Estimators} for a derivation).
\begin{figure*}[t]
\begin{center}
\includegraphics[width=0.95\textwidth]{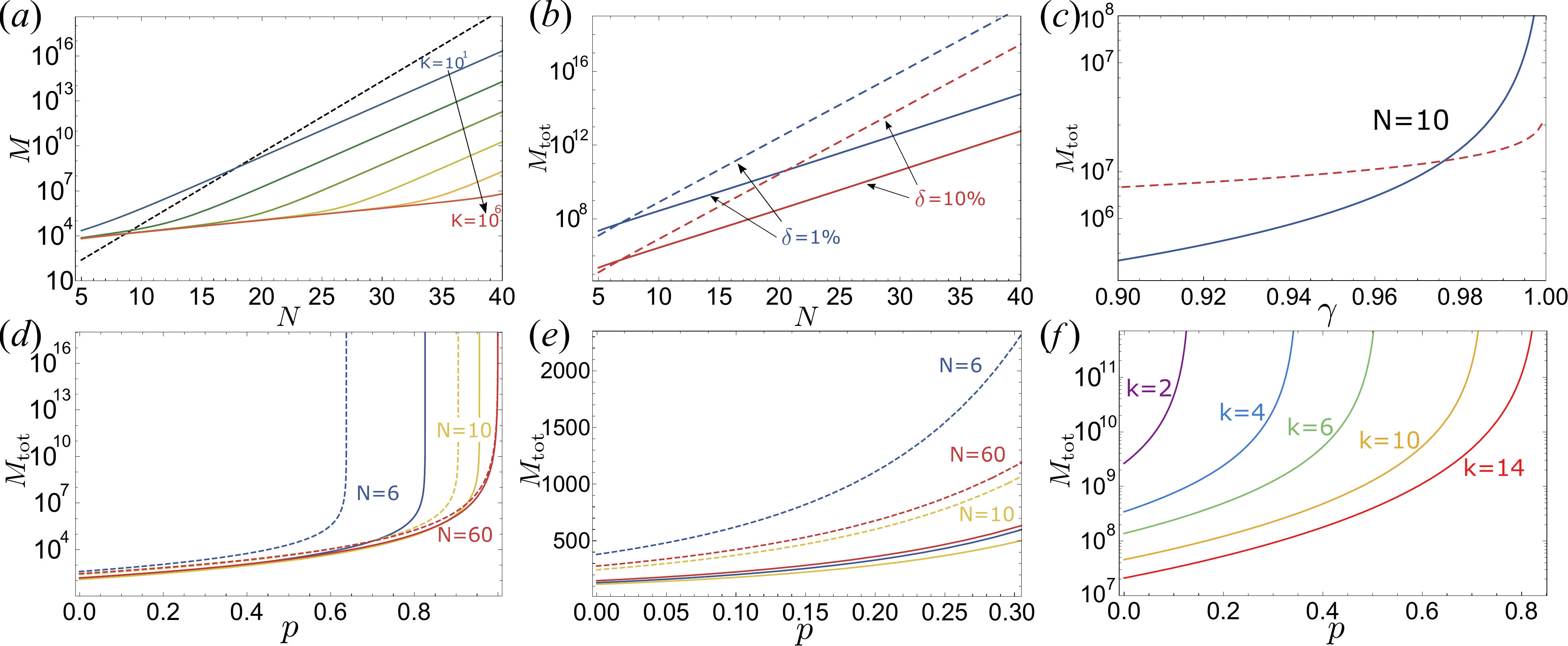}%
\end{center}
\caption{(a) Number $M$ of sampled measurement bases required to estimate $\mathcal R^{(2)}$ with an accuracy of at least $10\%$ and confidence  $\gamma=90\%$ as a function of the number of subsystems $N$ for $K=10,10^2,\ldots,10^{6}$ (solid curves from top to bottom), based on Chebyshev-Cantelli inequality. The black dashed line indicates the required measurement settings in order to exactly determine $\mathcal R^{(2)}$. (b) Total measurement budget $M_\text{tot}^{(\text{opt})}$  required for an estimation of $\mathcal R^{(2)}$ with accuracy $\delta=1\%$ (blue curve) and $10\%$ (red curve) as a function of $N$ obtained from Chebyshev-Cantelli (solid) and Bernstein (dashed) inequality. (c) $M_\text{tot}^{(\text{opt})}$ as a function of $\gamma$ for $N=10$ and $\delta=10\%$ obtained from Chebyshev-Cantelli (solid) and Bernstein (dashed) inequality.
(d,e) Measurement budget $M_\text{tot}^{(\text{opt})}$ obtained from Chebyshev-Cantelli inequality required to certify with confidence $\gamma=90\%$ that $\rho_\text{GHZ}^{(N)}(p)$ is  entangled (solid lines) or not in the $W$-class (dashed lines) for $N=6$ (blue), $N=10$ (yellow), and $N=60$ (red) qubits as a function of the $p$. (e) Zoom in of (d) for $0\leq p\leq 0.3$. (f) Same plot as in (d) but for the violation of the $k$-separability criteria (\ref{eq:ksepcrit}), with $k=2$ (violet, left), $4$ (blue), $6$ (green), $10$ (yellow) and $14$ (red, right), for $N=30$.
}
\label{fig_3}
\end{figure*}
\\
\indent
Hence, the precision of an estimation of the second moment is determined through Eqs.~(\ref{eq:Precision1}) and (\ref{eq:VarianceR2tilde}) and thus depends on the state under consideration. However, by bounding the variance~(\ref{eq:VarianceR2tilde}) from above we can consider a worst-case scenario and determine the required values of $M$ and $K$ in order to reach a precision of at least $\delta$ with confidence $\gamma$ (see App.~\ref{app:Estimators}).  To do so, we use the conjecture that the maximum of the fourth moment $\mathcal R^{(4)}$, for $N>4$, is attained by the $N$-qubit GHZ states. While this assumption is backed by numerical evidence we leave its proof for future investigations. 

In Fig.~\ref{fig_3}(a) we present the scaling of the required number of random measurement bases $M$ with the number of subsystems $N$ for different values of $K$. First, we note that the present statistical treatment allows for an improvement over the $3^N$ measurement settings that are required in order to evaluate the second moment exactly using a quantum design~\cite{tran1,tran2,MeMoments1,MeMoments2}, at the expense of a non-zero statistical error from the unitary sampling. Second, the required number of random measurement settings $M$ depends strongly on the chosen number of projective measurements per random unitary. More precisely, the curves in Fig.~\ref{fig_3}(a) scale as $\mathcal O (1.2^N)$ up to a threshold value that depends on $K$, beyond that the scaling with changes to $\mathcal O (2.25^N)$.

The minimum of  $M_\text{tot}=M\times K$ is reached for an optimal ratio between $M$ and $K$ which can be obtained analytically (see App.~\ref{app:MeasResourcesCantelli}) leading to $M_\text{tot}^{(\text{opt})}=M(K^{(\text{opt})})\times K^{(\text{opt})}$, as presented in Fig.~\ref{fig_3}(b). We thus find that the total measurement budget follows the overall scaling law $\mathcal O(1.5^N)$. Furthermore, while the required measurement resources increase slightly with higher precision, i.e., smaller $\delta$, the asymptotic scaling remains the same. As comparison, we present in the same figure the value  $M_\text{tot}^{(\text{opt})}$  obtained from the Bernstein inequality. The latter avoids the additional assumption about the upper bound on the variance (\ref{eq:VarianceR2tilde}) but scales worse with the system size. On the other hand, for fixed $N$, the scaling of  $M_\text{tot}^{(\text{opt})}$ with the confidence $\gamma$ is improved, as illustrated in Fig.~\ref{fig_3}(c). 

\section{Finite statistics entanglement characterization}
In order to certify the violation of  the $k$-separability bounds~(\ref{eq:ksepcrit}) one has to ensure that the statistical error $\delta$ of  $\tilde{\mathcal R}^{(2)}$ does not exceed the amount of the observed violation. This can be ensured by choosing the total number of measurements appropriately according to the previously discussed methods. Even more, since we aim to exclude the hypothesis that the state is, e.g., $k$-separable, we can improve our procedure by invoking upper bounds on the variances~(\ref{eq:VarianceR2tilde}) for $k$-separable states, respectively, instead of the overall upper bound used in Fig.~\ref{fig_3}(a-c). As this can only be done using the Chebyshev-Cantelli inequality we will focus on this approach in the following.
\\
\indent We demonstrate the above procedure using the state $\rho_\text{GHZ}^{(N)}(p)$ and first determine the total number of measurements $M_\text{tot}^{(\text{opt})}$ required  to certify that it is not fully-separable (i.e. $\mathcal R^{(2)} \leq 1/3^N$) and not in the class of $W$-states~\cite{tran1,tran2,MeMoments1,MeMoments2} (see Fig.~\ref{fig_3}(d,e) and also App.~\ref{app:BoundsMoments}). We find that already  moderate numbers of $M_\text{tot}^{(\text{opt})}\lesssim 2000$ are enough to certify their violation for up to $N=60$ qubits. 
Divergences displayed in Fig.~\ref{fig_3}(d) are due to the asymptotically decreasing difference between the true value of $\mathcal R^{(2)}$ and the respective bound of the targeted criterion.
A similar behavior is observed for the violation of different degrees of $k$-separability (see Fig.~\ref{fig_3}(f)). In this case $M_\text{tot}^{(\text{opt})}$ is generally on a higher level due to the increasing tightness of the bounds~(\ref{eq:ksepcrit}) for smaller $k$. \\
%
\section{Experimental implications}
Lastly, in order to demonstrate the applicability of our framework, we refer to recent experiments producing GHZ states with limited fidelity \cite{10qExpGHZ,20qExpGHZ2a,20qExpGHZ1,20qExpGHZ2}. For instance, in Ref.~\cite{20qExpGHZ1} a GHZ state of $11$ qubits was produced with fidelity $F\approx 0.75$. By applying our formalism we can thus show that the state contains at least $5$- or $7$-particle entanglement by performing in total of the order of $10^5$ or $10^6$ measurements, respectively (see App.~\ref{app:ExpGHZstates}). Note that these numbers are still moderate as compared to a full state tomography. Furthermore, we show that the $20$ qubit GHZ state of fidelity $F\approx 0.44$ (see Ref.~\cite{20qExpGHZ1}) contains at least $4$- or $5$-particle entanglement by performing in total of the order of $10^7$ measurements. We emphasize that such insights cannot be reached in terms of the fidelity, since fidelities up 
to $1/2$ can be reproduced by fully separable states.\\
%
\section{Conclusions}
We have discussed statistical methods allowing for the characterization of  multiparticle quantum systems based on randomized measurements. In particular, we presented novel criteria for the detection of different types of multiparticle correlations of $N$ qubit systems, including genuine multiparticle entanglement, based on the lowest non-vanishing moment only. Furthermore, we carried out a detailed  analysis of the involved statistical errors enabling an estimation of the  statistical significance of our methods. Lastly, we applied the developed framework in order to certify different types of multiparticle entanglement based on finite statistics and discussed applications to experiments in the noisy intermediate regime. \\

\begin{acknowledgements}
We thank Lukas Knips for discussions. AK acknowledges support by the Georg H. Endress foundation. SI acknowledges funding from the DAAD. NW acknowledges support by the QuantERA grant QuICHE and the German ministry of education and research (BMBF, grant no. 16KIS1119K). This work was supported by the Deutsche Forschungsgemeinschaft (DFG, German Research Foundation) - Projektnummern 447948357 and 440958198, and the ERC (Consolidator Grant 683107/TempoQ).
\end{acknowledgements}


\onecolumngrid

\vspace{8mm}
\begin{center}
{\large  APPENDIX}
\end{center}

\setcounter{secnumdepth}{2}
\setcounter{section}{0}
\section{\label{app:A}Moments of random correlations.} 
\subsection{Definition of moments}
For completeness we repeat below the main definitions presented at the beginning of the main text.  There we characterized a random measurement  through a set of random bases $\{(\ket{u_n^{(0)}}= U_n\ket{0_n},\ket{u_n^{(1)}}= U_n\ket{1_n})\}_{n=1,\ldots,N}$, with  $\{U_n\}_{n=1,\ldots,N}$ picked from the unitary group $\mathcal U(2)$ according to the Haar measure $\mu$. Further on, we  associate to each element $(\ket{u_n^{(0)}}= U_n\ket{0_n},\ket{u_n^{(1)}}= U_n\ket{1_n})$, with $n\in \{1,\ldots,N\}$, a direction $\boldsymbol u_n$ on the unit sphere $S^{2}$ with components $[\boldsymbol u_n]_i=\mathrm{tr}[\sigma_{\boldsymbol u_n}\sigma_i]$, with $i\in \{x,y,z\}$, and $\sigma_{\boldsymbol u_n}= U_n \sigma_z U_n^\dagger$ (see Fig.~1(a) of the main text). One such random measurement then enables us to retrieve the correlation functions:
\begin{align}
E^{(\alpha_1,\ldots,\alpha_{N'})}(\boldsymbol u_1,\ldots,\boldsymbol u_{N'})=\expec{ \sigma^{(\alpha_1)}_{\boldsymbol u_1} \ldots  \sigma^{(\alpha_{N'})}_{\boldsymbol u_{N'}}}_{\rho}.
\label{app:CorrelationFct}
\end{align}
where $ \sigma^{(\alpha_j)}_{\boldsymbol u_j}$ denotes the Pauli matrix $ \sigma_{\boldsymbol u_j}$ with support on the $\alpha_j$-th qubit, with $\alpha_j\in\{1,\ldots,N'\}$ and $N'\leq N$.

The correlation functions~(\ref{app:CorrelationFct}) provide a random snapshot of the correlation properties of the output state $\rho$. In order to get a more complete picture one has to consider the corresponding moments
\begin{align}
\mathcal R^{(t)}_{(\alpha_1,\ldots,\alpha_{N'})}= &\frac{1}{(4\pi)^{N'}} \int_{S^{2}} d\boldsymbol u_1\ldots \int_{S^{2}} d\boldsymbol u_{N'} \left[E^{(\alpha_1,\ldots,\alpha_{N'})}(\boldsymbol u_1,\ldots,\boldsymbol u_{N'})\right]^t, 
\label{app:RandomMoments}
\end{align}
where $t$ is a positive integer and $d\boldsymbol u_i=\sin{\theta_i} d\theta_i d\phi_i$ the uniform measure on the sphere $S^{2}$. Note that in the main text we focus solely on moments evaluated with full-correlation functions, i.e., correlation functions over all $N$ subsystems. In this case we drop the subscripts $(\alpha_1,\ldots,\alpha_{N'})$ and refer to it as the respective $N$-qubit moment.  As mentioned previously the moments (\ref{app:RandomMoments}) are by definition invariant under local unitary transformation, i.e., LU-invariant. Furthermore, due to the symmetries of the correlation functions (\ref{app:CorrelationFct})  with respect to a reflection on the local Bloch spheres, e.g. $E(\boldsymbol u_1,\ldots,-\boldsymbol u_i,\ldots, \boldsymbol u_N)=-E(\boldsymbol u_1,\ldots,\boldsymbol u_N)$, it is easy to see that all odd moments will vanish. We note that this is no longer true if one considers moments of quantum systems with larger local dimensionality~\cite{MeMoments2}.

In the following we will focus on the $N$-qubit moments and discuss several important properties. Eventually, we prove a criterion for the detection of genuine multipartite entanglement (GME) based on the second moment only. 

\subsection{Moments and spherical designs}
In Refs.~\cite{MeMoments1,MeMoments2} it was discussed that the integrals involved in the moments $\mathcal R^{(t)}$ can be replaced by an appropriate sum whenever an appropriate spherical $t$-design is known. In general, a spherical $t$-design in dimension three consist of a finite set of points $\{\boldsymbol u_{i}|i=1,\ldots,L^{(t)}\}\subset S^2$ fulfilling the property 
\begin{align}
\frac{1}{L^{(t)}} \sum_{k=1}^{L^{(t)}} P_{t'}(\boldsymbol u_k) = \frac{1}{4\pi}\int_{S^{2}}d\boldsymbol u \ P_{t'}(\boldsymbol u),
\label{app:tDesignDef}
\end{align} 
for all homogeneous polynomials $P_{t'}:S^2\rightarrow \mathbb R$, with $t'\leq t$. It thus suffices to resort to spherical $t$-designs as long as one is interested in calculating averages of polynomials of degree at most $t$ over the Bloch sphere $S^2$. This leads to the expression 
\begin{align}
\mathcal R^{(t)}
&=\frac{1}{(L^{(t)})^N} \sum_{k_1,\ldots,k_N=1}^{L^{(t)}} \expec{\sigma_{\boldsymbol u_{k_1}}\otimes \ldots \otimes \sigma_{\boldsymbol u_{k_N}}}^t.
\label{app:MomentSphericalDesign}
\end{align}  

Specifically, several concrete spherical designs on the $2$-sphere $S^2$ for $t$'s up to 20 and consisting of up to 100 elements are known (see for instance Ref.~\cite{ExamplesSphericalDesigns}). See Fig.~2 of Ref.~\cite{EURMe} for examples. Using these spherical designs the second moment can be expressed as follows
\begin{align}
\mathcal R^{(2)}&=\frac{1}{3^N} \sum_{i_1,\ldots,i_N=x,y,z} E(\boldsymbol e_{i_1}, \ldots  \boldsymbol e_{i_N})^2,\label{app:RandomMoment2}  
\end{align}
by summing over the three Pauli observables only. For higher order moments we need higher order designs, respectively. For instance, the fourth moment becomes
\begin{align}
\mathcal R^{(4)}&=\frac{1}{6^N} \sum_{i_1,\ldots,i_N=1}^{6} E(\boldsymbol v_{i_1}, \ldots  \boldsymbol v_{i_N})^4,
\label{app:RandomMoment4}
\end{align}
where the $\{\boldsymbol v_i|i=1,\ldots,L^{(5)}=12\}$ denotes the icosahedron $5$-design discussed in Ref.~\cite{EURMe}. Note that in Eqs.~(\ref{app:RandomMoment2}) and (\ref{app:RandomMoment4}) the number of summands is $L^{(t)}/2$ because for even $t$ one can drop the anti-parallel settings $-\boldsymbol e_i$ and $-\boldsymbol v_i$, respectively. 

\subsection{Bounds of the moments $\mathcal R^{(2)}$ and $\mathcal R^{(4)}$}\label{app:BoundsMoments}
Generally the moments (\ref{app:RandomMoments}) are upper and lower bounded depending on the class of quantum states under consideration. Furthermore, these bounds are in most cases dependent on the number of subsystems they are evaluated on. In this section we will summarize some known bounds thus yielding the basis for the novel GME-criterion introduced later on.

First, we consider the class $\mathcal B({(\mathbb C^2)}^{\otimes N})$ of all mixed $N$-qubit quantum states. In this case  all moments $\mathcal R^{(t)}$ are bounded from below by zero and equality is reached for the maximally mixed state. This is easy to see as all odd moments are zero and all even moments are positive. In contrast, it is much more difficult to determine tight upper bounds of the moments. Even only for the second moment $\mathcal R^{(2)}$ this problem has been solved only recently. In Refs.~\cite{tran2,JensR2max,NikolaiSectorLengths} it was shown that the maximum value of the second moment is reached for the $N$-qubit GHZ state $\ket{\text{GHZ}^{(N)}}=(\ket{0}^{\otimes N}+\ket{1}^{\otimes N})/\sqrt{2}$, yielding
\begin{align}
 \mathcal R^{(2)}_{\ket{\text{GHZ}^{(N)}}}=\frac{1}{3^N}\times \begin{cases}
2^{N-1}, &N\ \mathrm{odd}, \\
2^{N-1}+1,  &N\ \mathrm{even}.
\end{cases}
\label{app:GHZsts}
\end{align}
This insight was derived through the relation of the second moment to so-called sector lengths. The $N$-sector length is basically equal to Eq.~(\ref{app:RandomMoment2}) if one omits  the proportionality factor of $1/3^N$. Based on the framework of sector lengths it is also straightforward to extent the above values to the class of asymmetric GHZ states $a\ket{0}^{\otimes N}+b\ket{1}^{\otimes N}$, yielding 
\begin{align}
 \mathcal R^{(2)}_{\ket{\text{GHZ}^{(N)}_{a,b}}}=\frac{1}{3^N}\times \begin{cases}
(|a|^2-|b|^2)^2 +4|a|^2|b|^2  2^{N-1}, &N\ \mathrm{odd}, \\
4|a|^2|b|^2 2^{N-1} +1,  &N\ \mathrm{even}.
\end{cases}
\label{app:asymGHZsts}
\end{align}

Upper bounds of higher moments are unfortunately not known in general. However, we have numerical evidence that the upper bound of the fourth moment $\mathcal R^{(4)}$ is also reached for the GHZ state for which we find the values
\begin{align}
 \mathcal R^{(4)}_{\ket{\text{GHZ}^{(N)}}} =\frac{1}{15^N} \begin{cases}
3\times  8^{N-1}, &N\ \mathrm{odd}, \\
3\times  8^{N-1}+3^N+3\times 2^N,  &N\ \mathrm{even}.
\end{cases}
\label{app:R4GHZstate}
\end{align}
We did not find a state that has a larger fourth moment than the one of Eq.~(\ref{app:R4GHZstate}) except for the special case $N=4$. In this case the bi-separable state $\ket{\text{Bell}}\otimes\ket{\text{Bell}}$ reaches a larger value than the GHZ state. However, for an $N/2$-fold product of Bell states $\ket{\text{Bell}}^{\otimes (N/2)}$, with $N$ even, the fourth moment reads
\begin{align}
 \mathcal R^{(4)}_{\ket{\text{Bell}}^{\otimes \frac{N}{2}}} =\frac{1}{5^{(N/2)}}.
\label{app:R4BellProduct}
\end{align} 
Also, other product states like $\ket{\text{GHZ}_{\frac{N}{2}}}\otimes\ket{\text{GHZ}_\frac{N}{2}}$ have a smaller fourth moment than Eq. (\ref{app:R4GHZstate}), which is easy to check as the moments $\mathcal R^{(t)}$ factorize for product states.

Second, if we consider the class of separable states $\rho_\text{sep}=\sum_\alpha p_\alpha \rho^{(1)}_\alpha \otimes \ldots \otimes \rho^{(N)}_\alpha $ we can also derive upper bounds on the moments \cite{MeMoments1}. To do so, we simply exploit the convexity of the even moments which relies on the convexity of the monomials $x^t$ for even $t$. Furthermore, we know that for $N=1$ we find for all pure states $\mathcal R_{N=1}^{(2)}=\frac{1}{3}$ and $\mathcal R_{N=1}^{(4)}= \frac{1}{5}$. Hence, all in all we find the following bounds 
\begin{align}
\mathcal R^{(2)}\leq 1/3^N,\ \ \ \  \mathcal R^{(4)}\leq 1/5^N,
\label{app:RandomMoment4qu1}
\end{align}
for all separable $N$-qubit  states $\rho_\text{sep}$. 

Lastly, it has also been shown that one can derive upper bounds on the moments for different multipartite entanglement classes. For instance, in Ref.~\cite{MeMoments2} it was reported that the second moment is bounded from above by 
\begin{align}
\mathcal R^{(2)}\leq \frac{5-\frac{4}{N}}{3^N}=:\chi^{(N)}, 
\label{app:WclassBound}
\end{align}
for all states contained in the mixed $N$-qubit $W$-class. The latter is defined as $ \text{Conv}(\mathcal W^{(N)})$, where $\text{Conv}(\ldots)$ denotes the convex hull and $\mathcal W^{(N)}$ the pure $N$-qubit SLOCC (stochastic local operations and classical communication) class.

\subsection{Multiparticle entanglement criteria}\label{app:ProofMainCriteria}
In this section we prove the criteria introduced in Eq.~(3) of the main text. We start with the case $k=2$, i.e., the criterion allowing to detect genuinely multipartite entanglement.
To do so, let us first consider a pure biseparable state of $N$-qubits
\begin{align}
\rho_\text{bisep}=\rho_{N-k}\otimes \rho_k
\end{align}
with $k\in \{1,\ldots,N/2\}$, and calculate the maximum of its second moment
\begin{align}
\max_{\rho_\text{bisep}} \mathcal R^{(2)}_{\rho_\text{bisep}}&=\max_k \mathcal R^{(2)}_{\rho_{N-k}}\times \max_k \mathcal R^{(2)}_{\rho_k} \nonumber \\
&=\frac{1}{3^N}\left.\begin{cases}
2^{N-k-1}, &N-k\ \mathrm{odd} \\
2^{N-k-1}+1,  &N-k\ \mathrm{even}
\end{cases}\right\}\times \left.\begin{cases}
2^{k-1}, &k\ \mathrm{odd} \\
2^{k-1}+1,  &k\ \mathrm{even}
\end{cases}\right\},
\label{app:R2max1}
\end{align}
where we used that the maximum of an $m$-qubit second moment is attained for the respective $m$-qubit GHZ state (see Eq.~(\ref{app:GHZsts})). Further on, if $N$ is assumed to be even we find
\begin{align}
\max_{\rho_\text{bisep}} \mathcal R^{(2)}_{\rho_\text{bisep}}&=\frac{1}{3^N}\max_k{\left.\begin{cases} 2^{N-k-1}\times 2^{k-1},&k\ \mathrm{odd} \\ (2^{N-k-1}+1)\times (2^{k-1}+1),&k\ \mathrm{even} \end{cases} \right\}}
 \nonumber \\
&=\frac{1}{3^N}\max_k{\left.\begin{cases} 2^{N-2},&k\ \mathrm{odd} \\2^{N-2}+2^{N-k-1}+2^{k-1}+1,&k\ \mathrm{even}\end{cases}\right\}} \\
&=\frac{1}{3^N}\left[2^{N-2}+1+\max_k{(2^{N-k-1}+2^{k-1})} \right] \\
&=\frac{2^{N-3}+1}{3^{N-1}},
\end{align}
where we used that the function $f(k)=2^{N-k-1}+2^{k-1}$ is positive on the interval $[2,N/2]$ and thus takes its maximum at the boundary, i.e., for $k=2$ or $k=N-2$. Instead, if $N$ is odd we find
\begin{align}
\max_{\rho_\text{bisep}} \mathcal R^{(2)}_{\rho_\text{bisep}}&=\frac{1}{3^N}\max_k{\left.\begin{cases} (2^{N-k-1}+1)\times 2^{k-1},&k\ \mathrm{odd} \\ 2^{N-k-1}\times (2^{k-1}+1),&k\ \mathrm{even} \end{cases} \right\}}
 \nonumber \\
&=\frac{1}{3^N}\max_k{\left.\begin{cases} 2^{N-2}+2^{k-1},&k\ \mathrm{odd} \\ 2^{N-2}+2^{N-k-1},&k\ \mathrm{even}\end{cases}\right\}} \\
&=\frac{1}{3^N}\left[2^{N-2}+\max_k{\left. \begin{cases}2^{k-1},&k\ \mathrm{odd} \\ 2^{N-k-1},&k\ \mathrm{even}\end{cases}\right\}} \right] \\
&=\frac{1}{3^N}\left[ 2^{N-2}+\max_k{(2^{N-k-1})}\right] =\frac{2^{N-3}}{3^{N-1}},
\end{align}
where we used that $g(k)=2^{N-k-1}$ is positive in the interval $[2,N/2]$ and its maximum is reached for $k=2$. In summary, we thus proved that
\begin{align}
 \mathcal R^{(2)}\leq\frac{1}{3^{N-1}}\times \begin{cases}
2^{N-3}, &N\ \mathrm{odd}, \\
2^{N-3}+1,  &N\ \mathrm{even}.
\end{cases}
\label{app:GMEcrit}
\end{align}
for all biseparable states $\rho_\text{bisep}$. If we compare the bound (\ref{app:GMEcrit}) with the maximum value of the second moment (\ref{app:GHZsts}) we find that for odd number of qubits $2^{N-1}/3^N > 2^{N-3}/3^{N-1}$, for all $N$. For even number of qubits we have $(2^{N-1}+1)/3^N \geq (2^{N-3}+1)/3^{N-1}$, for all $N$, with equality iff $N=4$. Hence, the $N$-qubit second moment allows for the detection of genuine multipartite entanglement as long as $N\neq 4$. Furthermore, we note that the bounds in Eq.~(\ref{app:GMEcrit}) are saturated for the states $\ket{\text{Bell}}\otimes \ket{\text{GHZ}_{(N-2)}}$.

Further on, we show that for $k$-separable states $\mathcal R^{(2)}$ obeys the bounds
\begin{align}
 \mathcal R^{(2)}\leq\frac{1}{3^{N-k+1}}\times \begin{cases}
2^{N-(2k-1)}, &N\ \mathrm{odd}, \\
2^{N-(2k-1)}+1,  &N\ \mathrm{even},
\end{cases}
\label{app:ksepcrit}
\end{align}
with $k=2,\ldots,\left \lfloor{{(N-1)}/2}\right \rfloor $, a prove of which can be carried through the method of induction. Since we have proven Eq.~(\ref{app:ksepcrit}) in the case $k=2$, it remains the induction step, i.e., that the case $k+1$ follows from $k$. First, assume that $N$ is even and that for a $k$-separable state of $N-m$ qubits, denoted as $\rho_{N-m}$, the following holds
\begin{align}
 \mathcal R^{(2)}_{\rho_{N-m}}\leq\frac{1}{3^{N-m-k+1}}\times \begin{cases}
2^{N-m-(2k-1)}, &N-m\ \mathrm{odd}, \\
2^{N-m-(2k-1)}+1,  &N-m\ \mathrm{even}.
\label{app:R2ksepProof1}
\end{cases}
\end{align}
Now, we consider the maximum of the second moment of an $N$-qubit $(k+1)$-separable state
\begin{align}
\max_{\rho_{(k+1)\text{-sep}}} \mathcal R^{(2)}_{\rho_{(k+1)\text{-sep}}}&=\max_m \mathcal R^{(2)}_{\rho_{k\text{-sep},N-m}}\times \max_m \mathcal R^{(2)}_{\rho_{m}} 
\label{app:R2ksepProof2}
\end{align}
where $\rho_{k\text{-sep},x}$ denotes a $k$-separable state of $x$ qubits. Further on, we know by assumption that 
\begin{align}
\mathcal R^{(2)}_{\rho_{k\text{-sep},N-m}}\leq 
\frac{1}{3^{N-m-k+1}} (2^{N-m-(2k-1)}+\delta_{(N-m),\text{even}})
\label{app:R2ksepProof3}
\end{align}
and, according to Eq.~(\ref{app:GHZsts}), that 
\begin{align}
\mathcal R^{(2)}_{\rho_{m}}\leq\frac{1}{3^{m}} (2^{m-1}+\delta_{m,\text{even}}).
\label{app:R2ksepProof4}
\end{align}
Consequently, the RHS of Eq.~(\ref{app:R2ksepProof2}) becomes
\begin{align}
\max_{\rho_{(k+1)\text{-sep}}} \mathcal R^{(2)}_{\rho_{(k+1)\text{-sep}}}&=\max_m{\left\{\frac{1}{3^{N-m-k+1}}\times 
(2^{N-m-(2k-1)}+\delta_{(N-m),\text{even}})\times \frac{1}{3^{m}} (2^{m-1}+\delta_{m,\text{even}})\right\} }
\label{app:R2ksepProof5a}
\end{align}
which takes its maximum for even $m$ and thus leads to 
\begin{align}
\max_{\rho_{(k+1)\text{-sep}}} \mathcal R^{(2)}_{\rho_{(k+1)\text{-sep}}}&=\frac{1}{3^{N-k+1}}\times
\max_m{\{(2^{N-m-(2k-1)}+1)\times  (2^{m-1}+1)\}}\\
&=\frac{1}{3^{N-k+1}}\times 
\max_m{\{(2^{N-2k}+2^{N-m-(2k-1)}+2^{m-1}+1)\}},
\label{app:R2ksepProof5b}
\end{align}
where in the last two lines we assumed that $m$ is even. It thus remains to maximize Eq.~(\ref{app:R2ksepProof5b}) with respect to $m$. The $m$-dependent terms of Eq.~(\ref{app:R2ksepProof5b}) can be written as $g(M)=\frac{2^{N+1-2k}}{M}+\frac{M}{2}$, with $M:=2^m$, which is convex and thus attains its maximum at the boundary $M=2^m=2^2$, and thus for $m=2$. Altogether this leads to 
\begin{align}
 \mathcal R^{(2)}\leq\frac{2^{N-(2k-1)}}{3^{N-k+1}},
\label{app:R2ksepProof1}
\end{align}
for even $N$. An analogous calculation can be carried out for odd $N$. We note that the respective $k$-separability bounds~(\ref{app:ksepcrit}) are attained by the following class of pure $k$-separable states $\ket{\text{Bell}}^{\otimes (k-1)}\otimes \ket{\text{GHZ}_{N-2(k-1)}}$, which can be verified easily by explicitly evaluating the corresponding moment $\mathcal R^{(2)}$ for the respective states.

Similarly, we can formulate $k$-separable bounds of the fourth moment $\mathcal R^{(4)}$ which plays an important role for the determination of the measurement resources required in order to violate  Eq.~(\ref{app:ksepcrit}) with a given confidence. Based on the conjecture that for $N>4$ the $N$-qubit GHZ state maximizes the fourth moment $\mathcal R^{(4)}$ we can show that 
\begin{align}
 \mathcal R^{(4)}\leq\frac{1}{5^{k-1}} \mathcal R^{(4)}_{\ket{\text{GHZ}_{N-2(k-1)}}},
\label{app:R4ksepBounds}
\end{align}
using similar methods as in the proof of Eq.~(\ref{app:ksepcrit}). As for the second moment $\mathcal R^{(2)}$, the bound in Eq.~(\ref{app:R4ksepBounds}) is saturated for the $k$-separable pure states $\ket{\text{Bell}}^{\otimes (k-1)}\otimes \ket{\text{GHZ}_{N-2(k-1)}}$, with $k=2,\ldots,\left \lfloor{{(N-1)}/2}\right \rfloor $.

\subsection{Entanglement depth}\label{app:EntDepth}
Instead of bounding the moments $\mathcal R^{(t)}$ for $k$-separable states, one can derive bounds for the class of so-called $m$-producible states. The latter are characterized by the fact that they contain at least $m$-particle entanglement. More precisely, a pure state $\ket\Psi$ of $N$ particles is called producible by $m$-particle entanglement, i.e., $m$-producible, if it can be written as
\begin{align}
\ket\Psi=\ket{\phi_1}\otimes \ket{\phi_2}\otimes \ldots \otimes \ket{\phi_{\ell}}
\end{align}
where $\ket{\phi_i}$ are states of maximally $m$ particles and $\ell\geq N/m$. Hence, a state is called genuinely $m$-particle entangled if it is not producible by $(m-1)$-particle entanglement and thus has an entanglement depth of $m$. As for $k$-separable states, the above definition can be extended to mixed states by allowing for convex combinations of $m$-separable states. 

\begin{table}[t]
$N=11$:\begin{align}
\begin{array}{ V{2.5} c V{2.5} c  V{2.5} c V{2.5}}
\hlineB{2.5} m & \mathcal R^{(2)}_{m-\text{prod.}} & (k_1,k_2,\ldots,k_m) \\ 
\hlineB{2.5} 2 & \frac{1}{729} & (k_1= 1,k_2= 5) \\
\hline 3 & \frac{4}{2187} & (k_1= 0,k_2= 4,k_3= 1) \\
\hline 4 & \frac{4}{2187} & (k_1= 0,k_2= 0,k_3= 1,k_4= 2) \\
\hline 5 & \frac{16}{6561} & (k_1= 0,k_2= 1,k_3= 0,k_4= 1,k_5= 1) \\
\hline 6 & \frac{176}{59049} & (k_1= 0,k_2= 0,k_3= 0,k_4= 0,k_5= 1,k_6= 1) \\
\hline 7 & \frac{64}{19683} & (k_1= 0,k_2= 0,k_3= 0,k_4= 1,k_5= 0,k_6= 0,k_7= 1) \\
\hline 8 & \frac{64}{19683} & (k_1= 0,k_2= 0,k_3= 0,k_4= 1,k_5= 0,k_6= 0,k_7= 1,k_8= 0) \\
\hline 9 & \frac{256}{59049} & (k_1= 0,k_2= 1,k_3= 0,k_4= 0,k_5= 0,k_6= 0,k_7= 0,k_8= 0,k_9= 1) \\
\hline 10 & \frac{256}{59049} & (k_1= 0,k_2= 1,k_3= 0,k_4= 0,k_5= 0,k_6= 0,k_7= 0,k_8= 0,k_9= 1,k_{10}= 0) \\
 \hlineB{2.5}
\end{array}
\end{align}\\
$N=20$:\begin{align}
\begin{array}{ V{2.5} c V{2.5} c  V{2.5} c V{2.5}}
\hlineB{2.5} m & \mathcal R^{(2)}_{m-\text{prod.}} & (k_1,k_2,\ldots,k_m) \\ 
\hlineB{2.5} 2 & \frac{1}{59049} & (k_1= 0,k_2= 10) \\
\hline 3 & \frac{1}{59049} & (k_1= 0,k_2= 10,k_3= 0) \\
\hline 4 & \frac{1}{59049}& (k_1= 0,k_2= 0,k_3= 0,k_4= 5) \\
\hline 5 & \frac{65536}{3486784401} & (k_1= 0,k_2= 0,k_3= 0,k_4= 0,k_5= 4) \\
\hline 6 & \frac{65536}{3486784401} & (k_1= 0,k_2= 1,k_3= 0,k_4= 0,k_5= 0 ,k_6= 3) \\
\hline 7 & \frac{45056}{1162261467} & (k_1= 0,k_2= 0,k_3= 0,k_4= 0,k_5= 0,k_6= 1,k_7= 2) \\
\hline 8 & \frac{1849}{43046721} & (k_1= 0,k_2= 0,k_3= 0,k_4= 1,k_5= 0,k_6= 0,k_7= 0,k_8= 2) \\
\hline 9 & \frac{65536}{1162261467} & (k_1= 0,k_2= 1,k_3= 0,k_4= 0,k_5= 0,k_6= 0,k_7= 0,k_8= 0,k_9= 2) \\
\hline 10 & \frac{361}{4782969} & (k_1= 0,k_2= 0,k_3= 0,k_4= 0,k_5= 0,k_6= 0,k_7= 0,k_8= 0,k_9= 0,k_{10}= 2) \\
 \hlineB{2.5}
\end{array}
\end{align}
\caption{Numerical values of the $m$-producibility bounds, with $m=2,\ldots,10$, and the respective assignments $(k_1,k_2,\ldots,k_m)$ (see Eq.~(\ref{app:R2mprod2})) of the second moment $\mathcal R^{(2)}$ for $11$ (upper table) and $20$ (lower table) qubits.}
\label{tab:mprodbounds}
\end{table}

Based on the above definition we can proceed and derive bounds on the second moment $\mathcal R^{(2)}$ for $m$-producible states. To do so, we assume a pure $m$-separable states $\ket{\Psi_m}=\ket{\phi_1}\otimes \ket{\phi_2}\otimes \ldots \otimes \ket{\phi_{\ell}}$, where $\ket{\phi_j}$ consists of $m_j$ particles, for which we obtain
\begin{align}
\mathcal R^{(2)}_{\ket{\Psi_m}}&=\mathcal R^{(2)}_{\ket{\phi_1}}
\times
\mathcal R^{(2)}_{\ket{\phi_2}}\times \ldots \times \mathcal R^{(2)}_{\ket{\phi_\ell}}\\
&\leq  \mathcal R^{(2)}_{\ket{\text{GHZ}_{m_1}}}
\times
\mathcal R^{(2)}_{\ket{\text{GHZ}_{m_2}}}\times \ldots \times \mathcal R^{(2)}_{\ket{\text{GHZ}_{m_\ell}}}
\label{app:R2mprod1}
\end{align}
where we used that each $\mathcal R^{(2)}_{\ket{\phi_j}}$ is maximized by the respective $m_j$-qubit GHZ state. In order to maximize the RHS of Eq.~(\ref{app:R2mprod1}) we have to go simply through all possible assignments of $m_j$'s such that their sum equals to $N$. By rearranging terms we thus find:
\begin{align}
\mathcal R^{(2)}_{\ket{\Psi_m}}&\leq  \left(\mathcal R^{(2)}_{\ket{\text{GHZ}_{1}}}\right)^{k_1} \left(\mathcal R^{(2)}_{\ket{\text{GHZ}_{2}}}\right)^{k_2} \times \ldots \times \left(\mathcal R^{(2)}_{\ket{\text{GHZ}_{m}}}\right)^{k_m}
\label{app:R2mprod2}
\end{align}
with $\sum_{i=1}^m i k_i=N$, and where $\ket{\text{GHZ}_{1}}$ and $\ket{\text{GHZ}_{2}}$ refer to a single-qubit pure state and one of the Bell states, respectively. Finding the maximum of the RHS of Eq.~(\ref{app:R2mprod2}) for a given $N$ is thus a simple task. In Table~\ref{tab:mprodbounds} we give the $m$-producibility bounds of $\mathcal R^{(2)}$ for $N=11$ and $N=20$. Indeed, we find that for larger $m$'s the bounds often coincide with the $k$-separability bounds given in Eq.~(\ref{app:ksepcrit}), but can in general also differ from. While it might be possible to derive a general and concise formula, as for $k$-separable states (see Eq. (\ref{app:ksepcrit})), we leave this task for future investigations.

\subsection{Comparison with existing entanglement conditions}
In order to fairly compare our introduced multi-qubit entanglement conditions to existing ones we have to focus on those conditions which make comparable assumptions on the allowed measurement restrictions. 
In this respect, we first summarize entanglement conditions based on locally randomized measurements, which are invariant under local unitary transformations. That means that they do not require a shared reference frame between the involved parties and also relax the need of realizing a fixed local set of measurements. In this category, so far, there have been proposed bipartite entanglement conditions allowing to detect entanglement between two parts of a many-body system. The latter are based on the measurement of the entanglement entropy (see Ref.~\cite{ZollerScience}) or moments of the partially transposed density matrix (see Refs.~\cite{ElbenPRLmixedstate}). Entanglement detection with randomized measurements has also been discussed in Refs.~\cite{tran1,tran2}, however, there the focus was solely on detection non-full-separability. Hence, in the scenario of local randomized measurements our criteria are the first ones which allow for a detection of multi-qubit entanglement properties, i.e. non-$k$-separability, so there are no other criteria in the literature to which we can compare our criteria. \\
\indent Other entanglement conditions which make very similar assumptions on the measurement strategy are criteria which are invariant under local unitary transformation but are not based on randomized measurements, i.e., they require a fixed local set of measurement bases. One of the first works within this category is Ref.~\cite{BriegelLUinv} where local invariants have been introduced and showed how to used them for entanglement detection. Similar approaches have been discussed in Refs.~\cite{JulioMarkus,ZukowskiRefFrame1,ZukowskiRefFrame2}, which are based on norms of the correlation tensor. Finally, there is the concept of sector lengths (see Ref.~\cite{NikolaiSectorLengths}) which extends the ideas of these previous works and which served also as basis for the derivation of the present multi-qubit entanglement conditions. Note that any entanglement criterion involving the second moments $\mathcal R^{(2)}$ can be transformed into a criterion depending on sector lengths by using a unitary 2-design in order to evaluate the Haar averages contained in $\mathcal R^{(2)}$. However, we emphasize that by doing so the required number of measurements in order to evaluate the second moment $\mathcal R^{(2)}$ scales as  $3^N$ with the number $N$ of involved qubits. 

Finally, there are a number of criteria allowing for the detection of multiparticle entanglement properties based on non-$k$-separability which are, however, not invariant under local unitary transformations. Among the latter are approaches based on linear entanglement witnesses (e.g., \cite{ksepConditions1}) or more complicated nonlinear criteria (e.g., \cite{ksepConditions2}). While the latter criteria are often more favourable in terms of the required number of measurements for their evaluation, they usually rely on extensive prior information about the specific state under consideration. 

\section{Estimation of moments with finite statistics}
\label{app:EstimationMoments}

\subsection{General considerations}
The aim of this section is to provide several methods to estimate the statistical
error for the moments ${\mathcal R}^{(t)}$ if only a finite number of measurements 
has been performed. Before explaining the results in detail, we 
discuss the different strategies.

In general, in the case of finite statistics one tries to estimate the desired
quantities via (unbiased) estimators. These estimators coincide with the moments 
${\mathcal R}^{(t)}$ for infinite statistics, but in the finite case this is
is not necessarily true. Therefore, the aim is to derive deviation bounds, which give
upper bounds on the probability that the estimator deviates from the mean value
by a certain margin. 

In the present case, the experiment has a certain structure and includes 
different random processes. First, one draws $M$  different random unitaries 
and, for each unitary, a measurement is repeated $K$ times. This leads to 
several ways to derive deviation bounds:

\begin{itemize}
\item One may see the entire experiment as a single random variable, 
neglecting the structure outlined above. In this case, one can apply 
the Chebyshev-Cantelli inequality, which gives deviation bounds based on the 
variance of the entire random variable. This requires knowledge of 
the variance, which, as mentioned in the main text,  relies on some
assumptions on the maximal values of ${\mathcal R}^{(4)}.$ However, note that the estimates on the variance typically only hold if the measurements are properly implemented. 

\item 
In the iteration of drawing the $M$ different unitaries, one may see
any unitary as a different external parameter. In this sense, one has 
$M$ repetitions of a random variable, where the random variables are
independent, but differently distributed, due to the different unitary $U$.
For such scenarios there are tools to obtain deviation bounds, e.g., using
the Bernstein inequality. Here, no additional assumptions are required. 

\item 
One may see the $M$ different terms as independent and identically 
distributed variables. In each case, one draws a random unitary according
to the Haar measure and 
determines the correlation according to the rules of quantum mechanics. 
For this scenario, techniques using Chernoff bounds can be applied 
to derive estimation bounds in an analogous manner to the Hoeffding 
bound.  This gives typically the best bounds, but relies again on some 
assumptions on the maximal values of ${\mathcal R}^{(4)}.$
\end{itemize}

In the following subsections we will explain these approaches. First, we will 
discuss unbiased estimators for moments. Second, we present methods based on the 
Chebyshev-Cantelli inequality. The central result for this approach are the error bars given 
in Eqs.~(\ref{app:Precision1}, \ref{app:Precision1b}). Third, we explain an application of the Bernstein inequality, leading
to error bars with different scaling properties, see Eq.~(\ref{app-eq-bernstein}).
Finally, we consider the third approach. We show how Chernoff bounds 
can be used to derive various bounds for concrete situations. Depending
on the number $M$ of unitaries, this leads to different error estimates, 
see Eqs.~(\ref{app:Precisionfromchernoff}, \ref{app-eq-bernstein-2}). All the presented methods have their advantages and disadvantages, 
so it depends on the concrete experiment and its data, which method is favourable.

\subsection{Unbiased estimators and their variance}\label{app:Estimators}

As explained in the main text we denote individual outcomes of a single random measurement on $N$ qubits by $\{r_1,\ldots,r_{N}\}$, with $r_i=\pm1$, and the corresponding correlation sample as $X=\prod_{i=1}^{N}r_i$ (see Fig.~1(a) of the main text). Similarly, the correlation samples of subsets of $N'$ qubits are obtained  by  focusing on the respective outcomes $\{r_1^{(\alpha_1)},\ldots,r_{N'}^{(\alpha_{ N'})}\}$. Next, we  define the probability $P$ for obtaining the result $X=+1$, i.e., if an even number of the individual measurement outcomes $r_i$ resulted in $-1$. The corresponding unbiased estimator of $P$ can be defined as follows $\tilde P_1:=Y/K$, where $Y$ is a random variable distributed according to the binomial distribution with probability $P$ and $K$ trials. We thus  find $\mathbb E_\text{bi}(\tilde P_1)=P$, where $\mathbb E_\text{bi}(\ldots)$ denotes the average with respect to the binomial distribution. We note that similar methods have been employed also in Ref.~\cite{vermerschPRA} in the context of globally randomized measurement protocols.

Similarly, we can define unbiased estimator $\tilde P_k$ for the $k$-th powers of $P$ by making  the ansatz $\tilde P_k=\sum_{i=0}^k \alpha_i {(\tilde Y/K)}^i$ and enforcing the relation $\mathbb E(\tilde P_k)=P^k$. Hence, by using only the properties of the binomial distribution we can define unbiased estimators for arbitrary powers of $P$. For instance, we find 
\begin{align}
\tilde P_2&=\frac{\tilde P_1(K\tilde P_1-1)}{K-1}=\tilde P_1\times \frac{K\tilde P_1-1}{K-1},\\
\tilde P_3&=\frac{ \tilde P_1(K \tilde P_1-1)(K \tilde P_1-2)}{(K-1)(K-2)}=\tilde P_2\times\frac{K \tilde P_1-2}{K-2},\\
\tilde P_4&=\frac{\tilde P_1(K \tilde P_1-1)(K \tilde P_1-2)(K \tilde P_1-3)}{(K-1)(K-2)(K-3)}=\tilde P_3\times\frac{K \tilde P_1-3}{K-3},
\end{align}
which leads recursively to the following estimator for the $t$-th moment
\begin{align}
\tilde P_k&=\tilde P_{k-1}\times\frac{K \tilde P_1-(k-1)}{K-(k-1)}=\frac{\tilde P_1(K \tilde P_1-1)(K\tilde P_1-2)\ldots (K\tilde P_1-(k-1))}{(K-1)(K-2)\ldots (K-(k-1))}.
\label{app:UnbiasedEstPt}
\end{align}
Equation~(\ref{app:UnbiasedEstPt}) can be easily verified to be the unbiased estimator of $P^k$ by noting that the factorial moment of the binomial distribution reads $\mathbb E_\text{bi}[Y(Y-1)\ldots(Y-(k-1))]=K! P^t/(k-t)!$~\cite{FacMomentBinomDist}.
Further on, the unbiased estimators of the $t$-th powers of the correlations functions $E=2P-1$ can be obtained straightforwardly with formula
\begin{align}
\tilde E_t=(-1)^t\sum_{k=0}^t (-2)^k {t\choose k} \tilde P_k=(-1)^t\sum_{k=0}^t (-2)^k {t\choose k}\left[\frac{\tilde P_1(K \tilde P_1-1)(K\tilde P_1-2)\ldots (K\tilde P_1-(k-1))}{(K-1)(K-2)\ldots (K-(k-1))} \right],
\label{app:EstimatorEt}
\end{align}
which, in turn, allows us to define faithful estimators of the corresponding moments~(\ref{app:RandomMoments}):
\begin{align}
\tilde{\mathcal R}^{(t)}=\frac{1}{M}\sum_{i=1}^M  [\tilde E_t]_i.
\label{app:EstimatorRt}
\end{align}
We note that the the subscript $i$ refers to estimations of $\tilde E_t$ for different randomly sampled local bases, thus making the $[\tilde E_t]_i$ i.i.d. random variables. That is, we have $\mathbb E_{U}\mathbb E_\text{bi}\left[ (\tilde{\mathcal R}^{(t)})\right]=\mathcal R^{(t)}$, where $\mathbb E_U[...]$ denotes the average over local random unitaries.
We thus have provided a toolbox allowing for a statistical evaluation of the moment $\mathcal R^{(t)}$. In order to estimate the statistical errors of these evaluations we have to regard the variance of the respective unbiased estimators~(\ref{app:EstimatorRt}), that is
\begin{align}
\text{Var}\left(\tilde{\mathcal R}^{(t)}\right)=\mathbb E_{U}\mathbb E_\text{bi}\left[ (\tilde{\mathcal R}^{(t)}-\mathcal R^{(t)})^2\right]=\mathbb E_{U}\mathbb E_\text{bi}\left[ (\tilde{\mathcal R}^{(t)})^2\right]-\left({\mathcal R^{(t)}}\right)^2,
\end{align}
or by using Eq.~(\ref{app:EstimatorRt})
\begin{align}
\text{Var}\left(\tilde{\mathcal R}^{(t)}\right)=\frac{1}{M^2}\sum_{i=1}^M  \text{Var}\left([\tilde E_t]_i\right), 
\end{align}
since the $[\tilde E_t]_i$ are i.i.d. random variables. Hence, it suffices to evaluate the variance
\begin{align}
\text{Var}\left([\tilde E_t]_i\right)=\mathbb E_{U}\mathbb E_\text{bi}\left[\tilde E_t^2\right]-\left(\mathbb E_{U}\left[E^t\right]\right)^2=\mathbb E_{U}\mathbb E_\text{bi}\left[\tilde E_t^2\right]-\left(\mathcal R^{(t)}\right)^2. 
\end{align}
where $\mathbb E_{U}\mathbb E_\text{bi}\left[\tilde E_t^2\right]$ is in general a function of the moments $\mathcal R^{(t)}$. For instance, if we focus on the particular case $t=2$, we find 
\begin{align}
\text{Var}\left([\tilde E_2]_i\right)=A(K) \mathcal R^{(4)}+B(K)\mathcal R^{(2)}+C(K)-\left(\mathcal R^{(2)}\right)^2, 
\label{app-vare2}
\end{align}
with 
\begin{align}
A(K)&:=\frac{K}{K-1}-\frac{5}{K-1}+\frac{6}{(K-1) K}, \\
B(K)&:=\frac{4}{K-1}-\frac{8}{(K-1) K}, \\ 
C(K)&:=\frac{2}{K (K-1)},
\end{align}
which leads to 
\begin{align}
\text{Var}\left(\tilde{\mathcal R}^{(2)}\right)=\frac{1}{M} \left[A(K)\mathcal R^{(4)}+B(K)\mathcal R^{(2)}+C(K)-\left(\mathcal R^{(2)}\right)^2\right].
\label{app:VarianceR2tilde}
\end{align}
In order to arrive at the worst case error discussed in the main text we upper bound Eq.~(\ref{app:VarianceR2tilde}) by discarding the term $({\mathcal R^{(2)}})^2$ and using Eqs.~(\ref{app:GHZsts}) and (\ref{app:R4GHZstate}) which leads to 
\begin{align}
\text{Var}\left(\tilde{\mathcal R}^{(2)}\right)&\leq \frac{1}{M} \left[A(K)\mathcal R^{(4)}_{\ket{\text{GHZ}^{(N)}}}+B(K)\mathcal R^{(2)}_{\ket{\text{GHZ}^{(N)}}}+C(K)\right] \nonumber \\
&=\frac{1}{M} \left[A(K)\left.\begin{cases}
3\times  8^{N-1}/15^N, &N\ \mathrm{odd}, \nonumber \\
(3\times  8^{N-1}+3^N+3\times 2^N)/15^N,  &N\ \mathrm{even}.
\end{cases}\right\} \right. \\
&\phantom{=} \left. +B(K)\times \left.\begin{cases}
2^{N-1}/3^N, &N\ \mathrm{odd}, \\
(2^{N-1}+1)/3^N,  &N\ \mathrm{even}.
\end{cases}\right\}+C(K)\right].
\label{app:UpperBoundVarR2}
\end{align}
Hence, we found a state independent upper bound of the error on the estimator $\tilde{\mathcal R}^{(2)}$ which still involves a dependence on the number of subsystems $N$. The latter is important because the maxima of the moments tend to decrease with increasing $N$. 

\subsection{Estimating the deviation of $\mathcal R^{(2)}$ using the Chebyshev-Cantelli 
inequality }\label{app:MeasResourcesCantelli}

Using the variance bound derived in Eq.~(\ref{app:UpperBoundVarR2}) we can now derive a lower bound on the number of measurements $M \times  K$ that is required in order to estimate $\mathcal R^{(2)}$ with accuracy $\delta $ and confidence $\gamma$ (see Fig.~2(a) of the main text). To do so, we first regard the two-sided Chebyshev-Cantelli (see Ref.~\cite{SchmidtSpringer2010}) inequality for the random variable $\tilde{\mathcal R}^{(t)}$ yielding
\begin{align}
\text{Prob}[|\tilde{\mathcal R}^{(t)}-{\mathcal R}^{(t)}|\geq \delta]\leq \frac{2\text{Var}\left(\tilde{\mathcal R}^{(t)}\right)}{\text{Var}\left(\tilde{\mathcal R}^{(t)}\right)+\delta^2},
\label{app-eqcantelli}
\end{align}
which, by requiring that the confidence $1-\text{Prob}[|\tilde{\mathcal R}^{(t)}-{\mathcal R}^{(t)}|\geq \delta]$ of this estimation is at least $\gamma$, leads to the following minimal two-sided error bar that guarantees this confidence: 
\begin{align}
\delta_\text{err} &= \sqrt{\frac{1+\gamma }{1-\gamma } \text{Var}{\left(\tilde{\mathcal R}^{(t)}\right)}}. 
\label{app:Precision1}
\end{align}
Note that in case of the one-sided Chebyshef-Cantelli inequality (see Eq.~(\ref{eq:CantelliIneq1s})), Eq.~(\ref{app:Precision1}) becomes 
\begin{align}
\delta_\text{err} &= \sqrt{\frac{\gamma }{1-\gamma } \text{Var}{\left(\tilde{\mathcal R}^{(t)}\right)}},
\label{app:Precision1a}
\end{align}
which is slightly tighter and will be used in the next Sec.~\ref{app:ExpGHZstates} for the detection of multiparticle entanglement.

Furthermore, for the estimation of the second moment we can impose the variance bound of Eq.~(\ref{app:UpperBoundVarR2}), yielding
\begin{align}
\delta_\text{err} &\leq \sqrt{\frac{1+\gamma }{1-\gamma } \frac{1}{M} \left[A(K)\mathcal R^{(4)}_{\ket{\text{GHZ}^{(N)}}}+B(K)\mathcal R^{(2)}_{\ket{\text{GHZ}^{(N)}}}+C(K)\right]}.
\label{app:Precision1b}
\end{align}
Now, Eq.~(\ref{app:Precision1}) allows us to derive the required numbers of measurements $M$ and $K$ in order to reach a given error $\delta$. Since the size of the interval $[0,\mathcal R^{(2)}_{\ket{\text{GHZ}^{(N)}}}]$ depends on the number of subsystems $N$, we ask for a minimum relative error, i.e., a fraction of the length of the whole length interval. Hence, in order to achieve an estimation of the second moment with a relative error $\delta_\text{rel}$ and with a confidence $\gamma$ we require at least the following number of random measurement settings
\begin{align}
M(K)=\frac{\gamma +1}{\gamma -1}\frac{ 16\ 3^{N} \left((K-2) \left(-\left(2^{N}+2\right)\right)-3^{N}\right)-(K-3) (K-2) \left(\frac{3}{5}\right)^{N} \left(3\ 2^{N+3}+8\ 3^{N}+3\ 8^{N}\right)}{2  (K-1) K \left(2^{N}+2\right)^2 \delta _{\text{rel}}^2},
\label{app:NumberOfUs}
\end{align}
which is also presented in Fig.~3(a) of the main text. Furthermore, in order to determine the optimal number of projective measurements per random measurement setting we minimize $M(K)\times K$ (see also Fig.~\ref{fig_1}(left)) with respect to $K$ and with fixed number of parties $N$. To do so, we fix the desired confidence to $\gamma=90\%$ which leads to
\begin{align}
K_\text{opt}=1+\sqrt{2} \sqrt{1 - \frac{8\ 5^{N} \left(2^{N}-3^{N}+2\right)}{3\ 2^{N+3}+8\ 3^{N}+3\ 8^{N}}},
\label{app:Kopt}
\end{align}
which interestingly does no longer depend on the size of the error $\delta_\text{rel}$. In summary, Eqs.~(\ref{app:NumberOfUs}) and (\ref{app:Kopt}) fix the ratio between $M$ and $K$ (see Fig.~\ref{fig_1}(right)) and thus the total number of required measurement runs $M_\text{tot}=M\times K_\text{opt}$ as a function of the system size $N$ (see Fig.~3(b) of the main text). 
\begin{figure}[t!]
\begin{center}
\includegraphics[width=0.47\textwidth]{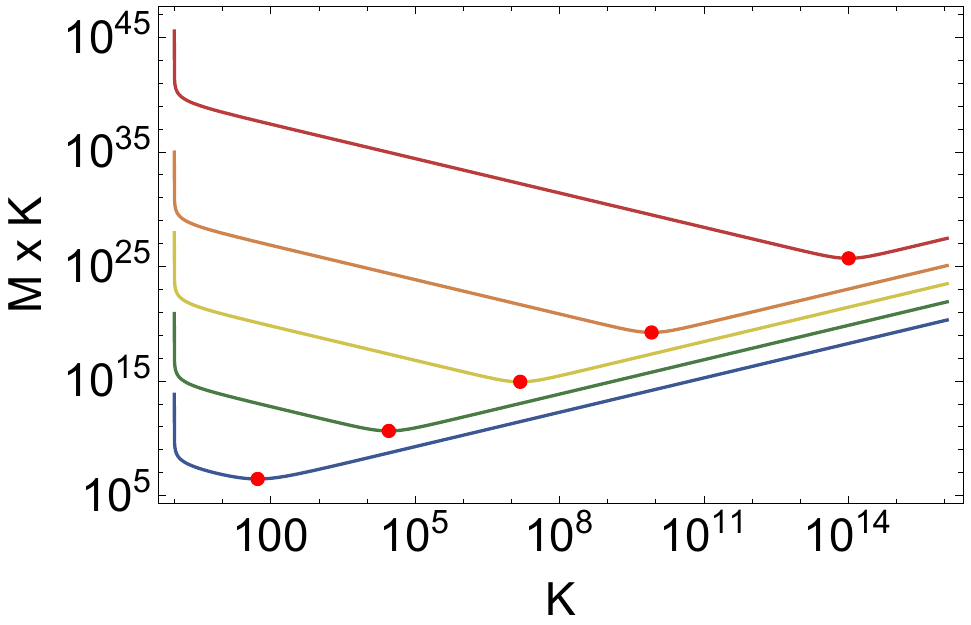}%
\includegraphics[width=0.47\textwidth]{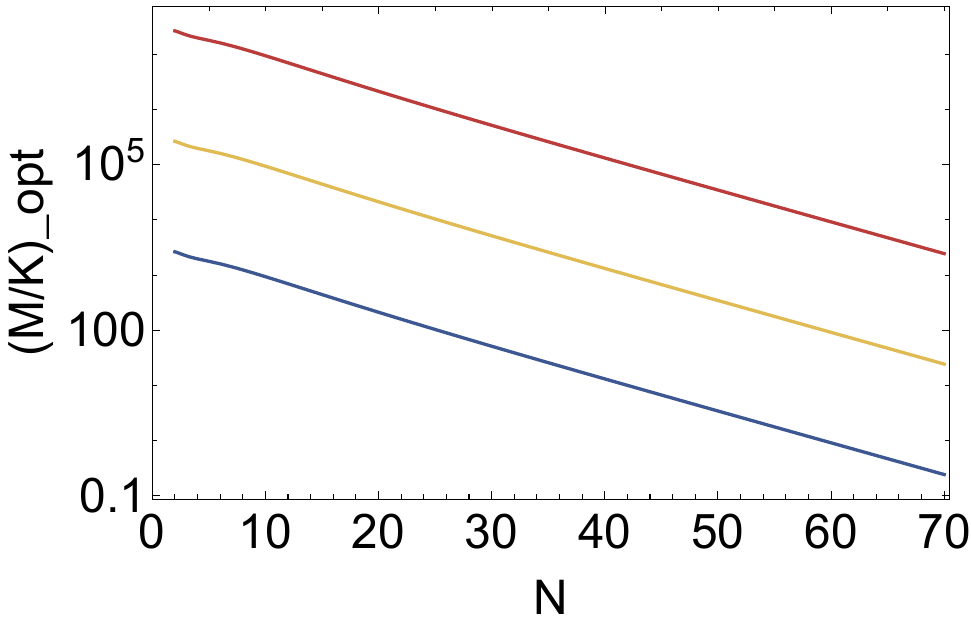}
\end{center}
\caption{Left: Total number of measurements $M_\text{tot}=M\times K$ as a funciton of $K$ for $N=10$, $30$, $50$, $70$ and $100$ qubits (from bottom to top) in order to estimate the second moment with a relative error of $10\%$ and with confidence $90\%$. Red dots indicate the position of the optimal value $M_\text{tot}^{(\text{opt})}$. Right: Optimal ratio $(M/K)^{(\text{opt})}$ as a function of the number of qubits $N$ for an estimation of the second moment with relative error $10\%$ (blue), $5\%$ (yellow), and $1\%$ (red) and confidence $90\%$.}
\label{fig_1}
\end{figure}

\subsection{Estimating the deviation of $\mathcal R^{(2)}$
using the Bernstein inequality}
\label{app:MeasResourcesBernstein}
In this section, we provide a different route to obtain deviation bounds based
on the Bernstein inequality. This inequality states the following \cite{bernsteinref}: 
Let $X_i \in[-c,c]$ be $M$ independent bounded random variables, let $\sigma_i^2$ 
be their variance and define $\sigma^2= \sum_{i=1}^M \sigma_i^2/M.$ Define the average
as $Z=\sum_{i=1}^M X_i/M.$ Then one has the bound
\begin{equation}
\text{Prob}[Z - \langle Z \rangle \geq \delta] \leq \exp\Big(-\frac{M\delta^2}{2\sigma^2+2c\delta/3M}\Big).
\label{app-eq-bernstein-0}
\end{equation}
In order to apply this, we focus on the moment ${\mathcal R}^{(2)}$ and 
we view the estimator in Eq.~(\ref{app:EstimatorRt}) as an average of $M$ 
independent random variables. Then, we have to study the estimator 
\begin{equation}
\tilde E_2 = 4 \tilde P_2 - 4 \tilde P_1  + 1 
=  \frac{4Y^2}{K(K-1)} - \frac{4 Y}{K} + 1
\end{equation}
as this corresponds to the variables $X_i$ in the Bernstein inequality.
Recall that $Y \in [0, K]$ is a binomially distributed variable with 
probability parameter $P$ coming from the $K$ trials. First, one can 
directly verify that $\tilde E_2 \in [-1/(K-1),1].$ 
This implies that we can take $c=1.$ Then, we have to compute the variance
of $\tilde E_2$. This is a fourth-order polynomial in $Y$ or $P$. 
Maximizing this over the admissible values of $P$ leads to the bound
\begin{equation}
\sigma_i^2 \leq \frac{2(K-1)}{K(2K-3)}    
\end{equation}
So, we obtain from the Bernstein inequality the two-sided bound
\begin{equation}
\text{Prob}[|\tilde{\mathcal R}^{(2)}-{\mathcal R}^{(2)}|
\geq \delta]
\leq
2 \exp\Big( - \frac{M \delta^2}{\frac{4(K-1)}{K(2K-3)}+\frac{2 \delta}{3 M}} \Big).
\end{equation}
Requiring that the confidence $1-\text{Prob}[|\tilde{\mathcal R}^{(2)}-{\mathcal R}^{(2)}|\geq \delta]$ of this estimation is at least $\gamma$, gives us the 
following minimal two-sided error bar that guarantees this confidence: 
\begin{align}
\delta_\text{err} =    \frac{|\log[(1-\gamma)/2]|}{3M^2}  
\Big( 
1 + \sqrt{ 1+\frac{36(K-1)M^3}{K(2K-3) \log[(1-\gamma)/2]} }
\Big) 
\label{app-eq-bernstein}
\end{align}
It is instructive to compare this with Eqs.~(\ref{app:Precision1}) and (\ref{app:Precision1b}). First, for fixed $\gamma$ and large $M$
both estimates show the same $\delta \sim 1/\sqrt{M}$ scaling. For fixed $M$ and 
$\gamma \rightarrow 1$, however, the Bernstein inequality scales significantly
better. If we set $\gamma = 1-\eta$ for small $\eta$, the error bars according
to the Chebyshev-Cantelli approach diverge as $\delta \sim \sqrt{1/\eta}$, while
the estimate according to the Bernstein inequality diverges only 
logarithmically as $\delta \sim \log(1/\eta)$.

\subsection{Estimating the deviation  of $\mathcal R^{(2)}$ using Chernoff bounds}
\label{app:MeasResourcesChernoff}
Finally, let us discuss approaches to derive error bars using 
Chernoff bounds and methods used for the proof of the Hoeffding inequality. In the end, it turns out that if $M$ is larger than
some (moderate) threshold, the bounds are strictly stronger 
than the ones from the Chebyshev-Cantelli inequality. The techniques for
deriving these bounds are a bit technical, although most of
these are standard tricks which can be found in various 
sources \cite{lecturenotes, taoblog}. 
Nevertheless, we present them here in some detail, 
as they can easily be modified to derive error bounds for 
some concrete experimental data.

{\bf Technical estimates for the exponential function.---}
We start with some technical estimates. 
First, we will need the bound
\begin{align}
\exp(x) \leq \exp\Big( \frac{1}{2\alpha}(x^2+\alpha^2)\Big)  
\label{app-exp-bound1}
\end{align}
for any $\alpha \geq 0.$ This follows trivially from the fact that
$(x-\alpha)^2 \geq 0.$
Second, we can use the exponential series to estimate for
$t \geq 0$ and $x \geq 0$
\begin{align}
\exp(tx) = 1 + tx + \frac{t^2 x^2}{2} ( 1 + \frac{tx}{3} + \frac{t^2x^2}{4 \cdot 3} + \dots )
\leq 1 + tx + \frac{t^2 x^2}{2} \exp(tx).
\label{app-exp-bound2}
\end{align}
This does not hold for negative $x$, but for this case we can just estimate
\begin{align}
\exp(tx) \leq 1 + tx + \frac{t^2 x^2}{2} \exp(|tx|). 
\label{app-exp-bound3}
\end{align}
Third, we can apply a similar trick for $x \geq 0$ to write
\begin{align}
\exp(t^2 x) = 1 + t^2 x \big( 1 + \frac{t^2 x}{2} + 
\frac{(t^2 x)^2}{3 \cdot 2} + \dots \big)    
\geq 
1 + (t^2 x) \exp\big(\frac{t^2 x}{2}\big).
\label{app-exp-bound4}
\end{align}

For the later application, we 
combine these estimates as follows. 
Assume that $\alpha \geq 0$ and $\beta \geq 0$. 
Then we have for all $\gamma \geq 0$
\begin{align}
1+t^2 \alpha \exp(t\beta) 
& \leq 1 + t^2 \alpha \exp\big[\frac{1}{2\gamma}(\beta^2t^2 + \gamma^2)\big],
\end{align}
which follows from Eq.~(\ref{app-exp-bound1}). If we now choose $\gamma$
such that
\begin{align}
\frac{1}{\gamma}(\beta^2 t^2 + \gamma^2) = t^2 \alpha^2,    
\label{app-gammacondition}
\end{align}
then we can apply Eq.~(\ref{app-exp-bound4}) to arrive at
\begin{align}
1+t^2 \alpha \exp(t\beta)  
\leq 
1 + t^2 \alpha \exp\big( \frac{t^2 \alpha^2}{2}\big) 
\leq 
\exp(t^2 \alpha).
\label{app-exp-bound5}
\end{align}
Clearly, the choice in Eq.~(\ref{app-gammacondition}) can not always be made, this
puts some restrictions on the parameters. In fact, Eq.~(\ref{app-gammacondition}) implies that
\begin{align}
\gamma = \frac{t^2 \alpha^2}{2} \pm \sqrt{\frac{t^4 \alpha^4}{4}-\beta^2 t^2},
\end{align}
which is only compatible with a positive and real $\gamma$ if 
\begin{align}
t^2 \alpha^4 \geq 4 \beta^2.   
\label{app-gammacondition2}
\end{align}

{\bf Applications to random variables.---}
Let us now consider a random variable $X \in [a,b]$ with $a \leq 0 \leq b$
and $\expec{X}=0.$ For that we have, using Eq.~(\ref{app-exp-bound3}),
\begin{align}
\mean{\exp(tX)} \leq \meanbig{1+ tX + \frac{t^2 X^2}{2} \exp(t\mu)}
= 1+ \frac{t^2 \expec{X^2}}{2} \exp(t\mu)
\end{align}
where we set $\mu = \max\{|a|, |b|\}.$ Using  Eq.~(\ref{app-exp-bound5})
we obtain the result
\begin{align}
\mean{\exp(tX)} \leq \exp\big(\frac{t^2 \expec{X^2}}{2}\big).
\label{app-lemmabound}
\end{align}
where $\expec{X^2}$ is also the variance $\text{Var}(X)$. 
Note that this estimate requires some relations between 
the parameters specified in Eq.~(\ref{app-gammacondition2}), 
this will be discussed at the end.

{\bf Deriving deviation bounds.---} Now we are in the position to apply
the preceding results to obtain deviation bounds. The general strategy
is the following \cite{lecturenotes}. Let $X_i$ be $M$ random variables and define 
$Z_i = \exp\big[t (X_i- \expec{X_i})\big]$. Assume that we have a bound
$\expec{Z_i} \leq \exp(C^2 t^2 /2).$ Then we can consider the variable
\begin{align}
Y = \frac{1}{M} \sum_{i=1}^M X_i    
\end{align}
and, using the Chernoff bound $\text{Prob}(R \geq a) \leq \expec{\exp(tR)} \exp(-ta)$
for a general random variable $R$ and 
all nonnegative $t$, we have 
for any $t \geq 0$
\begin{align}
\text{Prob}\big(Y-\expec{Y} \geq \delta \big) 
& \leq \meanbig{\exp\big[t(Y-\expec{Y})\big]} \exp(-t \delta)
\nonumber \\
& = \prod_{i=1}^M \meanbig{\exp\big[\frac{t(X_i-\expec{X_i})}{M}\big]} \exp(-t \delta)
\nonumber \\
& \leq \exp\big(\frac{C^2 t^2}{2M}-t \delta\big)
\nonumber \\
&
\!\!\!\!\!\!\!
\stackrel{t=\delta M / C^2}
{\leq}
\exp\big(-\delta^2 \frac{M}{2C^2}\big).
\label{app-niceestimate}
\end{align}

{\bf Application to randomized measurements.---} 
Now we have all the tools  for treating the physical situation under 
consideration. We first apply Eq.~(\ref{app-lemmabound}) to the random 
variable given by the unbiased estimator $X= \tilde{E}_2 - \expec{\tilde{E}_2}.$
This leads to 
\begin{align}
\meanbig{\exp\big[t(\tilde{E}_2 - \expec{\tilde{E}_2})\big]} 
\leq \exp\big(\frac{t^2 \text{Var}(\tilde{E}_2)}{2}\big),
\end{align}
where $\text{Var}(\tilde{E}_2)$ was already computed in Eq.~(\ref{app-vare2}).
Then, taking $Y= \tilde{\mathcal R}^{(2)}$ and using the method to derive 
deviation bounds we arrive at the main result, 
\begin{align}
\text{Prob}[|\tilde{\mathcal R}^{(2)}-{\mathcal R}^{(2)}|\geq \delta]
\leq 
2 \exp\Big(-\delta^2 \frac{M}{2 \text{Var}(\tilde{E}_2)}\Big).
\end{align}
Requiring again that the confidence $1-\text{Prob}[|\tilde{\mathcal R}^{(2)}-{\mathcal R}^{(2)}|\geq \delta]$ of this estimation is at least $\gamma$, and using the upper bound of the 
variance $\text{Var}(\tilde{E}_2)$ from Eq.~(\ref{app:UpperBoundVarR2})
gives us the following minimal two-sided error bar that guarantees this confidence: 
\begin{align}
\delta_\text{err} &=\sqrt{2|\log[(1-\gamma)/2]| \frac{1}{M} \left[A(K)\mathcal R^{(4)}_{\ket{\text{GHZ}^{(N)}}}+B(K)\mathcal R^{(2)}_{\ket{\text{GHZ}^{(N)}}}+C(K)\right]}.
\label{app:Precisionfromchernoff}
\end{align}
This is, up to the different scaling in the $\gamma$, the same error bar as in Eq.~(\ref{app:Precision1b}). In fact, one can directly check that for 
confidences $\gamma > 1/2$ one has 
$2|\log[(1-\gamma)/2]| \leq (1+\gamma)/(1-\gamma),$ so 
Eq.~(\ref{app:Precisionfromchernoff}) gives strictly better estimates 
than Eq.~(\ref{app:Precision1b}). For instance, for a confidence of 
$\gamma = 0.95$ (or $\gamma = 0.99$) the error bars from Eq.~(\ref{app:Precisionfromchernoff})
are by a factor $2.29$ (or $4.33$) smaller 
than the error bars from  Eq.~(\ref{app:Precision1b}).

Still, Eq.~(\ref{app:Precisionfromchernoff}) is only valid in a certain
parameter regime, and we finally have to discuss the conditions that
need to be fulfilled. Combining condition (\ref{app-gammacondition2})
with the choice of $t=\delta M / C^2$ in Eq.~(\ref{app-niceestimate})
leads, after a short calculation, to
\begin{align}
M \geq \frac{8 \mu}{ \delta \text{Var}(\tilde{E}_2)}.
\label{app-mbound}
\end{align}
This sets a minimal number $M$ of random unitaries  that need to be performed in order to make the error estimate valid. Note that here one can also use an upper bound on
the variance. In the previous calculations, the constant
$C$ was taken to be the variance $\text{Var}(\tilde{E}_2),$
but, of course it is valid for any number larger than that. Using an
upper bound on the variance leads to larger error bars as in 
Eq.~(\ref{app:Precisionfromchernoff}), but also to smaller values 
of $M$, where the deviation bound starts to be valid. 

Let us estimate the minimal $M$ for some scenarios. 
We have that $\tilde E_2 - \expec{\tilde E_2}  \in [-1-1/(K-1),1],$
so we take $\mu = 1+1/(K-1).$ Then, using the upper bound from Eq.~(\ref{app:UpperBoundVarR2}) we find for $N=5$ and $K=10$ 
the bound $M \geq 159.6/\delta$, for $N=10$ and $K=10$
the bound is $M \geq 345.3/\delta$ and for $N=5$ and $K=100$ 
one needs $M \geq 440.3/\delta$. The dependence on $K$ can be understood
 as follows: If $K$ is large, this results independently of $M$ in very small error bars in
 Eq.~(\ref{app:Precisionfromchernoff}) as the
 variance decreases with $K$. Naturally, also 
 the required $M$ has to increase in order to justify 
 small error bars. Still, the entire approach can be used
 for arbitrary $M$ and $K$, as described in the following.

First, it should be noted that the above theory can also be easily modified to work for smaller values of $M$. Indeed if Eq.~(\ref{app-mbound}) does not hold one can just define a constant
$C'= 8\mu / \delta M > \text{Var}(\tilde{E}_2)$ and use it
in the derivation of the deviation bound. This will give slightly increased error bars, where $\text{Var}(\tilde{E}_2)$ is replaced by the larger value $C'$.

Finally, this also gives a constructive way to compute an 
error bar for a given fixed $M$ and $K$ and given confidence $\gamma$. First, one can consider the error bar given in Eq.~(\ref{app:Precisionfromchernoff}) and checks
whether for the resulting $\delta = \delta_\text{err}$ 
and the used upper bound on the variance $\text{Var}(\tilde{E}_2)$ the condition 
in Eq.~(\ref{app-mbound}) holds. If this is the case, then one has a valid error bar. If this is not the case, one can increase the upper bound of the variance in  Eq.~(\ref{app:Precisionfromchernoff})  by 
a factor $\eta > 1$. Then, $\delta_\text{err}$ will increase by a factor
of $\sqrt{\eta}.$ Consequently, the condition Eq.~(\ref{app-mbound}) 
on $M$ will become significantly weaker, as a factor of 
$\sqrt{\eta} \times \eta$ arises in the denominator. In fact, the 
minimal $\eta$ can directly be computed from this, giving the 
increased error bar $\sqrt{\eta} \delta_\text{err}.$

{\bf A second application of the Bernstein inequality.---} 
Finally, we would like to mention that the Bernstein inequality can
also be applied to the considered scenario. That is, we consider $M$
independent and identically distributed variables with a variance 
$\text{Var}(\tilde{E}_2).$ Then, in Eq.~(\ref{app-eq-bernstein-0})
we have $\sigma^2 = \text{Var}(\tilde{E}_2)$ and $c = \mu = 1+1/(K-1).$
This leads, in analogy to Eq.~(\ref{app-eq-bernstein}), to
\begin{align}
\delta_\text{err} =    \frac{\mu|\log[(1-\gamma)/2]|}{3M^2}  
\Big( 
1 + \sqrt{ 1+\frac{18 M^3 \text{Var}(\tilde{E}_2)}{\mu^2\log[(1-\gamma)/2]} }
\Big)
>
\sqrt{2 |\log[(1-\gamma)/2]| \frac{1}{M}\text{Var}(\tilde{E}_2)}.
\label{app-eq-bernstein-2}
\end{align}
So, this approach delivers slightly worse error bars in comparison with 
Eq.~(\ref{app:Precisionfromchernoff}), but it has the advantage to be 
applicable to any $M$~\cite{TobiasNauck}.  Finally, note that application of Eq.~(\ref{app-eq-bernstein-2}) usually requires the assumption of an 
upper bound on $\text{Var}(\tilde{E}_2)$, which is not needed in 
Eq.~(\ref{app-eq-bernstein}).

\section{Characterizing multiparticle entanglement with randomized measurements}

\subsection{Entanglement properties of the noisy GHZ state $\rho_\text{GHZ}^{(N)}(p)$}\label{app:ExpGHZstates}
Using the methods introduced in the previous section we can determine the measurement resources required for the detection of different types of multiparticle entanglement given a predefined  confidence $\gamma$ (see Fig.~3(d-f) of the main text). For the remainder of this discussion we will resort to the method based on the Chebyshev-Cantelli inequality discussed in Sec.~\ref{app:MeasResourcesCantelli}. The reason for this is, on the one hand that this methods yielded an overall better performance for the estimation of the second moment $\mathcal R^{2}$, as compared to the method based on the Bernstein inequality (see  Sec.~\ref{app:MeasResourcesBernstein}). On the other hand, we found that the method based on Chernoff bounds, discussed in Sec.~\ref{app:MeasResourcesChernoff}, yields a slightly better result than the Chebysheff-Cantelli but it involves some technical considerations about the allowed number of measurements (see the discussion around Eq.~\ref{app-mbound}) which makes it slightly less practical to apply. Furthermore, the methods of Sec.~\ref{app:MeasResourcesCantelli} and \ref{app:MeasResourcesChernoff} have the additional advantage that they involve the upper bound~(\ref{app:UpperBoundVarR2}) on the variance of the estimator $\tilde{\mathcal R}^{(2)}$ which can be further refined in the context of the detection of $k$-separable states (see below for further details). 

Hence, we start from the one-sided Chebyshev-Cantelli inequality 
\begin{align}
\text{Prob}[\tilde{\mathcal R}^{(2)}-{\mathcal R}^{(2)}\geq \delta]\leq \frac{\text{Var}\left(\tilde{\mathcal R}^{(2)}\right)}{\text{Var}\left(\tilde{\mathcal R}^{(2)}\right)+\delta^2}.
\label{eq:CantelliIneq1s}
\end{align}
The one-sided version suffices in the scenario of entanglement detection, since we only have to show that $\mathcal R^{(2)}$ is larger than the bounds of the criteria~(\ref{app:ksepcrit}). Also, since we want to rule out the hypothesis that the state belongs to a certain class of separable states, we can further invoke the respective upper bounds of the second and fourth moment in order to obtain an upper bound on the variance $\text{Var}\left(\tilde{\mathcal R}^{(2)}\right)$. In doing so we further improve the required measurement resources in comparison to the overall worst-case scenario for an estimation of $\mathcal R^{(2)}$ considered in Sec.~\ref{app:EstimationMoments}. 

For instance, for the detection of non-$k$-separability we use Eqs.~(\ref{app:ksepcrit}) and (\ref{app:R4ksepBounds}) in order to upper bound the variance in Eq.~(\ref{eq:CantelliIneq1s}). Furthermore, we set the confidence $\gamma=90\%$ and the accuracy $\delta =\mathcal R_\text{GHZ}^{(2)}(p,N)-\max{\mathcal R^{(2)}_{\rho_{k\text{-sep}}}}=(1-p)^2  \mathcal R^{(2)}_{\ket{\text{GHZ}^{(N)}}}-\max{\mathcal R^{(2)}_{\rho_{k\text{-sep}}}}$, where $\max{\mathcal R^{(2)}_{\rho_{k\text{-sep}}}}$ denotes the RHS of Eq.~(\ref{app:ksepcrit}), such that the state $\rho_\text{GHZ}^{(N)}(p)$ violates the respective $k$-separability bound. Now, it remains to determine the optimal total number of measurements $M^{(\text{opt})}_\text{tot}$, as demonstrated in Sec.~\ref{app:MeasResourcesCantelli}. The results of this calculation are presented in Fig.~4 of the main text for the detection of different degrees of multiparticle entanglement, i.e., violation of $k$-separability, and the discrimination of $W$-class entanglement, according to criterion~(\ref{app:WclassBound}), for different values of the noise parameter $p$ and the number of qubits $N$.

Lastly, we give the precise measurement numbers required for the entanglement depth detection discussed in the last section of the main text. There we considered GHZ states of $N=11$ and $N=20$ qubits, with fidelities taken according to recent experiments reported in Ref.~\cite{20qExpGHZ1}. The number of required measurements, in order to prove with a confidence of $90\%$ that a GHZ state of $N=11$ qubits and fidelity $F=0.76$ has an entanglement depth of $5$ or $7$, is:
\begin{align}
    M_{\text{tot}}^{(\text{opt})}&=(M\times K)^{(\text{opt})}=3685 \times 125 \approx 4.60625 \times 10^5, \label{app:NumMeas1}\\
    M_{\text{tot}}^{(\text{opt})}&=(M\times K)^{(\text{opt})}=571082 \times 105 \approx 5.996361 \times 10^7,
    \label{app:NumMeas2}
\end{align}
respectively. Furthermore, to prove entanglement depth $4$ or $5$, respectively, of a GHZ state of $N=20$ qubits and fidelity $F=0.44$ one requires 
\begin{align}
    M_{\text{tot}}^{(\text{opt})}&=(M\times K)^{(\text{opt})}=11062 \times 4875 \approx 5.392725 \times 10^7, \label{app:NumMeas3}\\
    M_{\text{tot}}^{(\text{opt})}&=(M\times K)^{(\text{opt})}=18752 \times 4420 \approx 8.288384 \times 10^7.
    \label{app:NumMeas4}
\end{align}
As explained previously the above numbers are based on the analytically determined optimal ratio between $M$ and $K$ for the respective type of entanglement under consideration. 

We note that in Ref.~\cite{20qExpGHZ1} the number of performed measurements used to estimate the lower bounds on the fidelity is given by $(2N+2)\times 16384$. While the latter numbers are only one to two orders of magnitude smaller than those reported in Eqs.~(\ref{app:NumMeas1})-(\ref{app:NumMeas4}), the  corresponding experimental procedure, based on a quantum sensing circuit, is considerably more involved. Furthermore, our method requires stabilization of the experiment only for the time of performing $K$ measurements of a single randomly chosen measurement setting which might be of a general advantage in experiments based on Rydberg atom arrays or superconducting qubits. Lastly, we emphasize that our methods reveal information about the entanglement structure of the states also in regimes of low fidelities where fidelity-based witnesses cannot give any insight. This is due to the fact that fidelity values of $F\leq 1/2$ can always be reproduced by fully separable states.

\subsection{Other local noise sources}\label{app:OtherNoiseSources}
The global depolarization model yields the following behaviour of the second
moment: Given a state $\rho$ which yields a second moment $\mathcal R^{(2)}_{\rho}$ then the corresponding
value of $\mathcal R^{(2)}$ after global depolarisation (with probability $p$) is given by $\mathcal R^{(2)}_{\rho,\text{depol}}
=\mathcal R^{(2)}_\rho (1-p)^2$. Hence, the global depolarisation leads to a quadratic decay of the
second moment with the depolarisation probability p. This behaviour was used in the
manuscript in order to investigate the detection of GHZ state entanglement subject to
depolarisation noise.

In case of a local depolarisation model the situation is similar but with one major
difference. The second moment of a state $\rho$ subject to local depolarisation decays as
$\mathcal R^{(2)}_{\rho,\text{depol}}
=\mathcal R^{(2)}_\rho (1-q)^{2 L}$, where $L$ denotes the number of applications of the
local depolarisation noise channel and $q$ the local depolarisation probability. Hence, from
the latter it seems that our methods are more vulnerable to local noise, however, we
emphasise that the global and local depolarisation probabilities $p$ and $q$ cannot be directly
compared.

To illustrate the impact of local depolarisation we will consider the following practical
example. The two gates of the quantum computing platform discussed in the section
experimental considerations (see Ref. [37] of the main text) have average error rates of
approximately $q=0.01$
the two-qubit error) which translates roughly into local depolarisation probabilities of the
same size. Hence, the global depolarisation probability of a $N$-qubit GHZ state, which
requires the application of $1$ single-qubit gate and $N-1$ two-qubit gates, can be estimated
by adding the respective local depolarisation rates as follows $p=1-(1-0.01)^{(N-1)}$. In case
of $N=20$ this yields approximately $p=0.82$ which can be used in order to apply the methods
presented in our manuscript.


\end{document}